\documentclass[twocolumn,aps,floatfix,citeautoscript,superscriptaddress,cite]{revtex4} %
\usepackage{graphicx}
\usepackage{color}
\usepackage{amsmath}
\usepackage[caption=false,labelformat=empty,position=top,font=bf]{subfig}

\begin{document}

\title{Tunable magnon topology in monolayer CrI$_\mathbf{3}$ under external stimuli}

\author{Maarten Soenen}
\affiliation{Department of Physics \& NANOlab Center of Excellence, University of Antwerp, Groenenborgerlaan 171, B-2020 Antwerp, Belgium}

\author{Milorad V. Milo\v{s}evi\'c}
\email{milorad.milosevic@uantwerpen.be}
\affiliation{Department of Physics \& NANOlab Center of Excellence, University of Antwerp, Groenenborgerlaan 171, B-2020 Antwerp, Belgium}
\affiliation{Instituto de Física, Universidade Federal de Mato Grosso, Cuiabá, Mato Grosso 78060-900, Brazil}

\begin{abstract}
Two-dimensional (2D) honeycomb ferromagnets, such as monolayer chromium-trihalides, are predicted to behave as topological magnon insulators - characterized by an insulating bulk and topologically protected edge states, giving rise to a thermal magnon Hall effect. Here we report the behavior of the topological magnons in monolayer CrI$_3$ under external stimuli, including biaxial and uniaxial strain, electric gating, as well as in-plane and out-of-plane magnetic field, revealing that one can thereby tailor the magnetic states as well as the size and the topology of the magnonic bandgap. These findings broaden the perspective of using 2D magnetic materials to design topological magnonic devices.
\end{abstract}

\date{\today}

\maketitle

\section{Introduction}\label{sec:intro}
Last years have witnessed a surge of research interest in the field of topological magnonics \cite{mcclarty2022}, driven by experimental findings in diverse materials, including pyrochloric \cite{onose2010,ideue2012}, perovskite \cite{ideue2012}, hexagonal \cite{chen2018,chen2021,cai2021,zhu2021}, and kagome \cite{chisnell2015,hirschberger2015} structures. These materials, exhibiting non-trivial magnon topology, manifest a bandgap in their bulk magnon bands, and topologically protected edge states, leading to a thermal magnon Hall effect. Materials with such properties are known as \emph{topological magnon insulators} (TMIs), drawing an analogy to their electronic counterparts. Several recent studies proposed that the properties of TMIs can be exploited to construct a variety of topological magnonic devices, including waveguides, spin wave diodes, beam splitters, and interferometers \cite{mook2015,rukriegel2018,wang2018}. In order to realize such devices experimentally, and optimize their performance, it is important to understand how to control and tune the magnon topology of a given material. 

CrI$_3$, the archetypal two-dimensional (2D) magnetic material, has recently gained significant attention in the field of magnonics, as it possesses highly-tunable magnon modes in the terahertz frequency range \cite{jin2018,cenker2021,menezes2022}, exhibiting potential for fast and energy efficient data processing applications \cite{roadmap}. In addition, several theoretical studies \cite{aguilera2020,costa2020,soenen2023} suggested that 2D CrI$_3$ may be a promising candidate to host topological magnons, leading to the possibility of realizing TMIs in the monolayer limit.  

For a material to posses topological magnons, at least two conditions need to be fulfilled: \emph{(i)} the spin-wave dispersion has a non-zero bandgap, in CrI$_3$, the magnonic bandgap  emerges from the Dzyaloshinskii-Moriya interaction (DMI) and the Kitaev interaction, which both originate from the spin-orbit coupling (SOC) \cite{chen2018,chen2021,aguilera2020,costa2020,soenen2023,kim2022,owerre2016,zhang2021}, and are prone to tuning through mechanical and electric stimuli \cite{menezes2022,zheng2018,bacaksiz2021,sabani-arxiv}, offering opportunities to tailor the magnonic bandgap of CrI$_3$; \emph{(ii)} the effective time-reversal symmetry is be broken, e.g. due to the spontaneous magnetization of a material or an externally applied magnetic field. However, even when these two conditions are fulfilled, the magnon topology is known to depend on intrinsic material properties like sublattice symmetry \cite{soenen2023,kim2022}, stacking order \cite{soenen2023}, or magnetic configuration \cite{soenen2023}. In the present work, we demonstrate that the magnon topology of monolayer CrI$_3$ can also be actively tuned using external stimuli such as biaxial and uniaxial strain, and applied electric or magnetic fields. 

The paper is organized as follows. In Sec. \ref{sec:method}, we detail the computational methodology used in this work. We provide a detailed description of the Heisenberg Hamiltonian that models the magnetic interactions in CrI$_3$ and explain how the magnetic parameters of this Hamiltonian are obtained from first-principles calculations. Further, we discuss how we determine the ground-state spin configuration and the magnonic dispersion for a given set of magnetic parameters, and define the Chern number which will be used to classify the topology of the magnons. In Sec. \ref{sec:results}, we first briefly revise the magnetic and magnonic properties of a pristine CrI$_3$ monolayer in Sec. \ref{sec:pristine}, and then show how these properties are influenced by biaxial and uniaxial strain, electric gating and an external magnetic field in Secs. \ref{sec:biax_strain} -- \ref{sec:mag_field} respectively. For biaxial strain, we report magnetic phase transitions between the out-of-plane ferromagnetic (FM) spin polarization of the pristine case, to in-plane FM and N\'eel antiferromagnetic (AFM) states. Further, we reveal that one can tune both the size and the topology of the magnonic bandgap using biaxial and uniaxial strain, and out-of-plane and in-plane magnetic fields, unveiling topological phase transitions between phases with opposite Chern numbers, and between phases with trivial and non-trivial topology. Furthermore, we investigate the effect of an applied electric field on monolayer CrI$_3$, albeit causing rather small quantitative changes to the magnonic dispersion. Sec. \ref{sec:conclu} summarizes our findings and discusses further research opportunities within the field. Additionally, in the Appendix, we report results for the spin-wave dispersion of the pristine CrBr$_3$ and CrCl$_3$ monolayers (belonging to the same Cr-trihalide family), revealing the presence of a very small topological bandgap for the former and a Dirac point for the latter. 

\section{Methodology and theory}\label{sec:method}
The magnetic interactions in CrI$_3$ are modeled by a Heisenberg spin Hamiltonian: 
\begin{equation}\label{eq:ham}
 \hat{\mathcal{H}}=\frac{1}{2}\sum_{i , j} \mathbf{\hat{S}}_i \mathcal{J}_{ij} \mathbf{\hat{S}}_j + \sum_i \mathbf{\hat{S}}_i \mathcal{A}_{ii}\mathbf{\hat{S}}_i + \mu_\mathrm{B}\sum_i\mathbf{B}\cdot g_i \mathbf{\hat{S}}_i,
\end{equation} 
in which $\mathcal{J}_{ij}$ and $\mathcal{A}_{ii}$ are the exchange and single-ion anisotropy (SIA) matrices respectively, and the spins $\mathbf{\hat{S}}_i = (\hat{S}_{i}^{x},\hat{S}_{i}^{y},\hat{S}_{i}^{z})$ are expressed in Cartesian coordinates. We consider a spin of S = 3/2, since the chromium atoms show a magnetic moment of $\mu = 3\mu_B$. The exchange term of the Hamiltonian can be decomposed in separate contributions as:
\begin{align}
    \hat{\mathcal{H}}_\mathrm{ex} = \frac{1}{2}\sum_{i , j} \left[J_{ij} \mathbf{\hat{S}}_i \cdot \mathbf{\hat{S}}_j + K_{ij} \hat{S}_i^{\gamma}  \hat{S}_j^{\gamma} + \mathbf{D}_{ij} (\mathbf{\hat{S}}_i \times \mathbf{\hat{S}}_j)\right], \nonumber
\end{align}
where the spins $\mathbf{\hat{S}}_i = (\hat{S}_{i}^{\alpha},\hat{S}_{i}^{\beta},\hat{S}_{i}^{\gamma})$ are now considered in the local eigenbases $\{\alpha,\beta,\gamma\}$ that diagonalize the symmetric parts of the exchange matrices. In these bases, we define the isotropic exchange constant as $J_{ij} = (J_{ij}^\alpha + J_{ij}^\beta)/2$, and the Kitaev constant as $K_{ij} = J_{ij}^{\gamma} - J_{ij}$. The components of the DMI are calculated in the Cartesian basis from the off-diagonal elements of the exchange matrix as $D^x_{ij} = \frac{1}{2}(\mathcal{J}^{yz}_{ij}-\mathcal{J}^{zy}_{ij})$, $D^y_{ij} = \frac{1}{2}(\mathcal{J}^{zx}_{ij}-\mathcal{J}^{xz}_{ij})$ and $D^z_{ij} = \frac{1}{2}(\mathcal{J}^{xy}_{ij}-\mathcal{J}^{yx}_{ij})$ \cite{xiang2013,sabani2020}. Notice that the sign of the DMI vector depends on the hopping direction of the considered spin pair, since $\mathbf{D}_{ij} = \nu_{ij}|\mathbf{D}_{ij}|$ with $\nu_{ij} = -\nu_{ji} = \pm 1$. Due to the rotational symmetry of CrI$_3$, it's sufficient to calculate only one element of the SIA matrix, namely $\mathcal{A}_{ii}^{zz}$, all the other matrix elements are redundant \cite{xiang2013,sabani2020}. However, breaking of this symmetry, e.g. by applying uniaxial strain, requires a calculation of the full SIA-matrix. The effect of an external magnetic field is accounted for by the inclusion of a term due to the Zeeman interaction in the Heisenberg Hamiltonian, where $g_i\approx 2$ is the $g$-factor, and $\mu_\mathrm{B}$ is the Bohr magneton. The Heisenberg Hamiltonian is parameterized from first principles using the four-state energy mapping (4SM) methodology \cite{xiang2013,sabani2020} and density functional theory (DFT). A detailed discussion on the implementation of the needed DFT calculations is included in Appendix \ref{app:dft}. Tables containing all the calculated exchange and SIA parameters can be found in the supplemental material \cite{sup_mat}.

The ground-state spin configurations and magnonic dispersion of monolayer CrI$_3$ are both calculated using \textsc{spinw} \cite{spinw}. The ground state spin configurations are obtained using the magnetic structure optimizer functionalities that are implemented in \textsc{spinw}. These functions can reach (local) energy minima by iteratively rotating the spins starting from an initial random spin configuration. The minimization was repeated for varying supercell sizes to exclude its influence on the final result. To obtain the spin-wave dispersion, we assume linear spin-wave theory and perform a numerical diagonalization of the Heisenberg Hamiltonian in reciprocal space.

To classify the topology of magnons, we calculate the Chern number, which is a topological invariant with an integer value that is defined for the $n$th band as:
\begin{align}\label{eq:chern}
    \mathcal{C}_n = \frac{1}{2\pi i}\int_\mathrm{BZ} \Omega_{n\mathbf{k}}\ d^2k,
\end{align}
where the integration is performed over the first Brillouin zone (BZ), and the Berry curvature is equal to:
\begin{align}
    \Omega_{n\mathbf{k}} = i \sum_{n^{\prime} \neq n} \frac{\langle n\left|\partial_\mathbf{k} \right. \hat{\mathcal{H}}_\mathbf{k} \left| n^{\prime} \right. \rangle \langle n^{\prime}\left|\partial_\mathbf{k} \right. \hat{\mathcal{H}}_\mathbf{k} \left| n\right. \rangle}{\left(\lambda_{n\mathbf{k}}-\lambda_{n^{\prime}\mathbf{k}}\right)^2},
\end{align}
with $\lambda_{n\mathbf{k}}$ and $|n\rangle$ respectively the eigenvalues and eigenvectors of the Heisenberg Hamiltonian $\hat{\mathcal{H}}_\mathbf{k}$ in reciprocal space. For systems that are gapless or show a trivial bandgap, the Chern numbers vanish. In this work, we calculate Chern numbers according to the link-variable approach as detailed in Ref.~\cite{fukui2005} for a discretized BZ.

\begin{figure}[b!]
\includegraphics[width=0.85\linewidth]{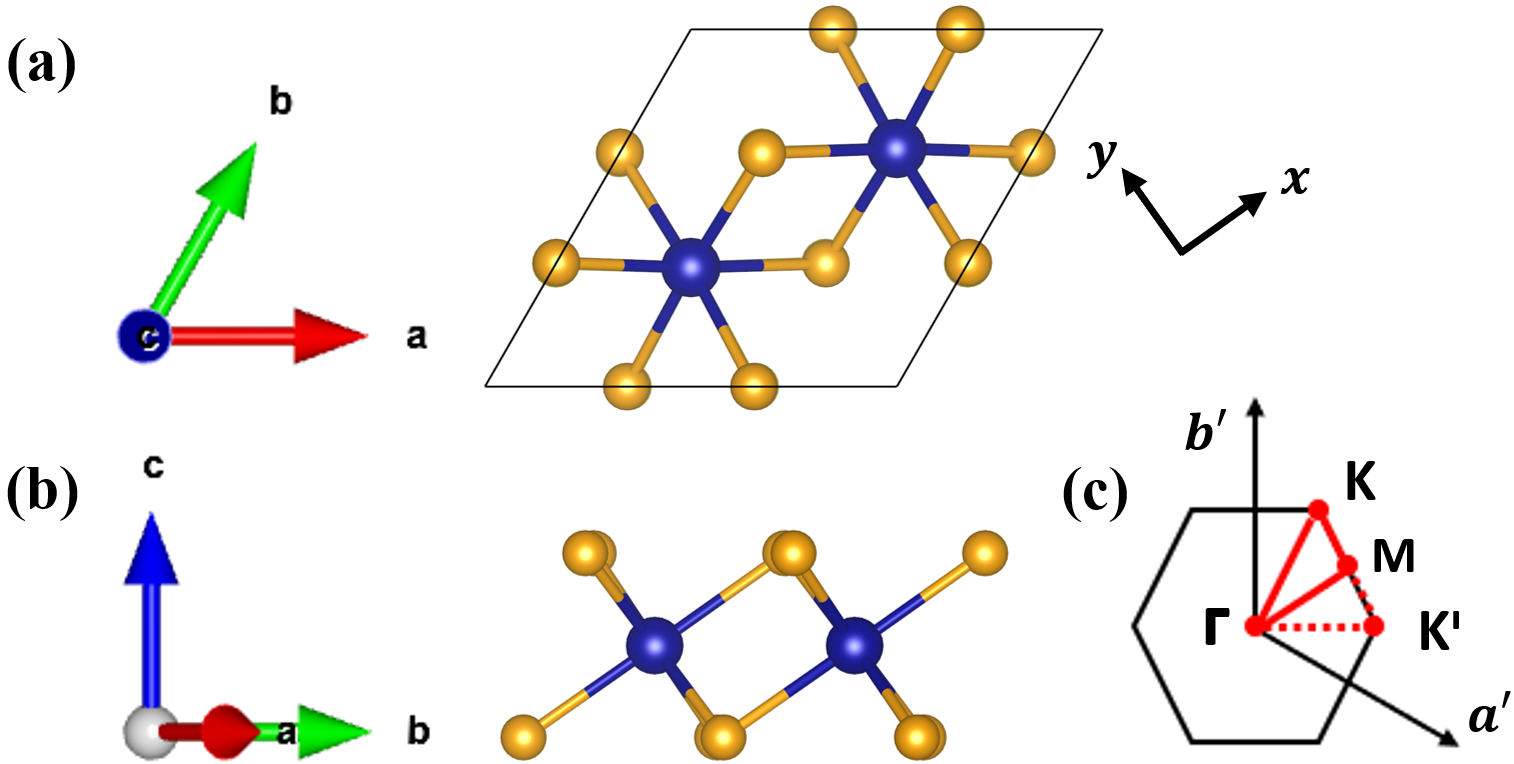}
\caption{\label{fig:structure} Top view (a) and side view (b) of the crystal structure of monolayer CrI$_3$. The chromium and iodine atoms are depicted with dark blue and orange spheres respectively. The unit cell is marked with a solid black line. The crystal structure was drawn using \textsc{vesta} \cite{VESTA}. Panel (c) depicts the corresponding first BZ and high-symmetry points for 2D systems with a hexagonal lattice.}
\end{figure}

\section{Results}\label{sec:results}
\subsection{Pristine monolayer}\label{sec:pristine}
Before discussing the tuning of magnons using external stimuli, we first review the results for a pristine CrI$_3$ monolayer \cite{soenen2023}. Figs. \ref{fig:structure}(a) and \ref{fig:structure}(b) depict the crystal structure of monolayer CrI$_3$. The chromium atoms form a honeycomb lattice and are octahedrally coordinated with six iodine atoms. Each unit cell contains two chromium atoms, which are connected through two $\approx 90^\circ$ Cr-I-Cr bonds. Structural relaxation using DFT, yields an in-plane lattice constant of $a$ = 6.919 \AA. Fig.~\ref{fig:structure}(c) depicts the first BZ and high-symmetry points of a 2D hexagonal lattice. Note that, in principle, the $K$ and $K'$ points are inequivalent, however, if the sublattice symmetry is upheld, the dispersion will be identical at both points. 

Monolayer CrI$_3$ has a FM ground state, caused by isotropic exchange constants of $J_{\mathrm{NN}}$ = -4.35 meV and $J_{\mathrm{NNN}}$ = -0.74 meV for respectively the nearest-neighbor (NN) and next-nearest-neighbor (NNN) interactions. The anisotropy in the exchange interaction is captured by the Kitaev constants, which are equal to $K_{\mathrm{NN}}$ = 1.49 meV and $K_{\mathrm{NNN}}$ = 0.17 meV. This exchange anisotropy, together with a SIA parameter of $\langle \mathcal{A}^{zz}_{ii}\rangle$ = -0.08 meV, causes the spins to prefer an out-of-plane orientation. Due to the presence of an inversion center between NN spins (i.e. preserved inversion symmetry), the resulting DMI will be zero. However, there is no inversion symmetry between NNN spins, resulting in a weak yet non-zero DMI of $|\mathbf{D}_{\mathrm{NNN}}|$ = 0.06 meV.

The calculated magnetic parameters for the pristine CrI$_3$ monolayer yield the spin-wave dispersion depicted as the blue curve in Fig.~\ref{fig:biaxial_strain_dispersion}(a). The lower energy `acoustic' branch and the higher energy `optical' branch, correspond to an in-phase and an out-of-phase precession of the spin sublattices respectively. At the $\Gamma$ point, a Goldstone gap of $\Delta_\Gamma = 0.43$ meV opens in the dispersion due to the material's magnetic anisotropy. At the $K$ point, there is a small bandgap of $\Delta_K = 0.15$ meV, which is caused by a combination of the NNN DMI and the NN Kitaev interaction. This bandgap turns out to be topologically non-trivial, yielding Chern numbers of $\mathcal{C}_n = \pm 1$ for the upper and lower bands, respectively. The origin of the topology is attributed to the breaking of time-reversal symmetry due to the spontaneous magnetization of CrI$_3$. Note that, although the calculated bandgap is of the same order of magnitude as other theoretically calculated values for monolayer CrI$_3$ \cite{olsen2021,gorni-arxiv}, it is significantly smaller than the experimentally observed value of $\approx$ 2.8 meV for bulk CrI$_3$ reported in Ref.~\cite{chen2021}. Although this issue is still open to debate, a recent study suggests that the magnon-phonon coupling, which we do not account for in our model, is the mechanism responsible for this discrepancy \cite{delugas-arxiv}.

\subsection{Effect of biaxial strain}\label{sec:biax_strain}
In this section, we discuss to which extent the spin states and the magnonic properties of monolayer CrI$_3$ can be tuned by applied biaxial strain. 

\begin{figure}[tp!]
\centering
\subfloat[\raggedright(a)]{\includegraphics[width=.9\linewidth]{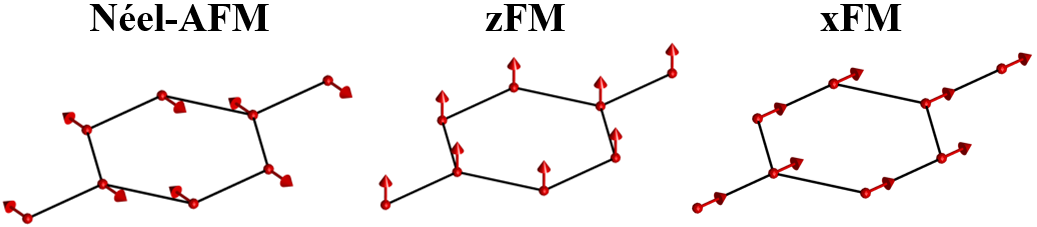}} \\
\subfloat[\raggedright(b)]{\includegraphics[width=.9\linewidth]{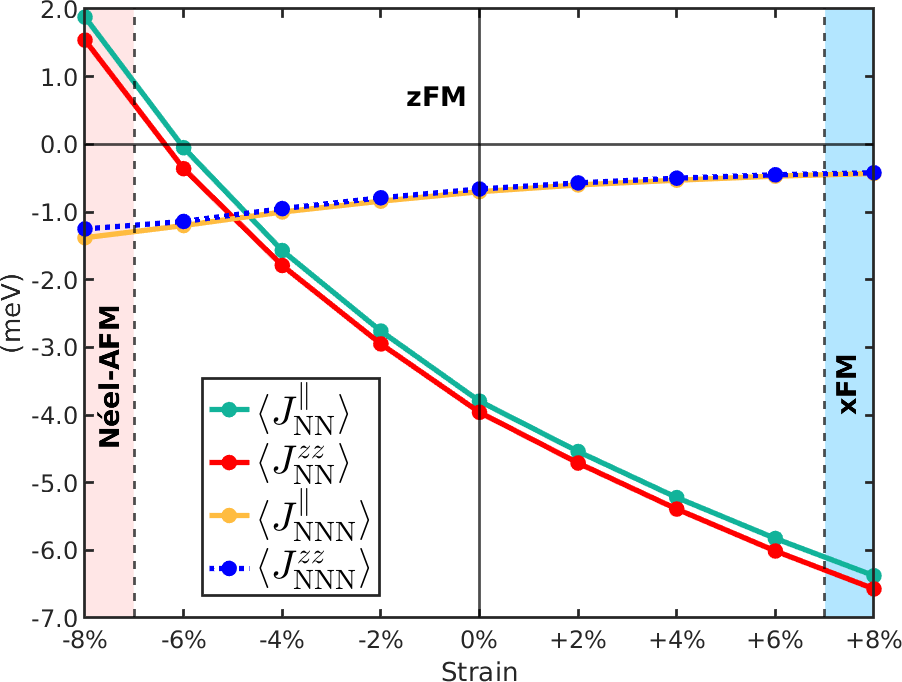}} \\
\subfloat[\raggedright(c)]{\includegraphics[width=.9\linewidth]{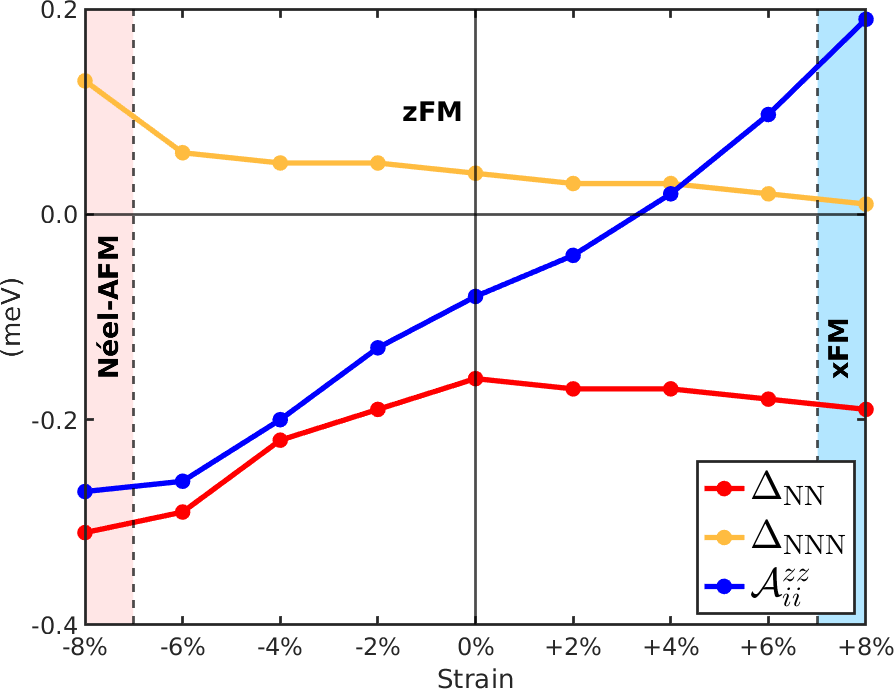}}
\caption{\label{fig:biaxial_strain_parameters} Effect of biaxial strain on the magnetic properties of monolayer CrI$_3$. (a) Spin configurations of a biaxially strained CrI$_3$ monolayer. (b) Evolution of the in-plane and out-of-plane exchange constants as a function of the applied strain. (c) Dependence of the out-of-plane exchange anisotropy and the SIA on the applied strain. All parameters are considered in the Cartesian basis. Dotted lines mark magnetic phase transitions between the spin configurations depicted in (a).}
\end{figure}

\begin{figure*}[tp!]
\centering
\hspace*{\fill}
\subfloat[\raggedright(a)]{\includegraphics[width=.45\textwidth]{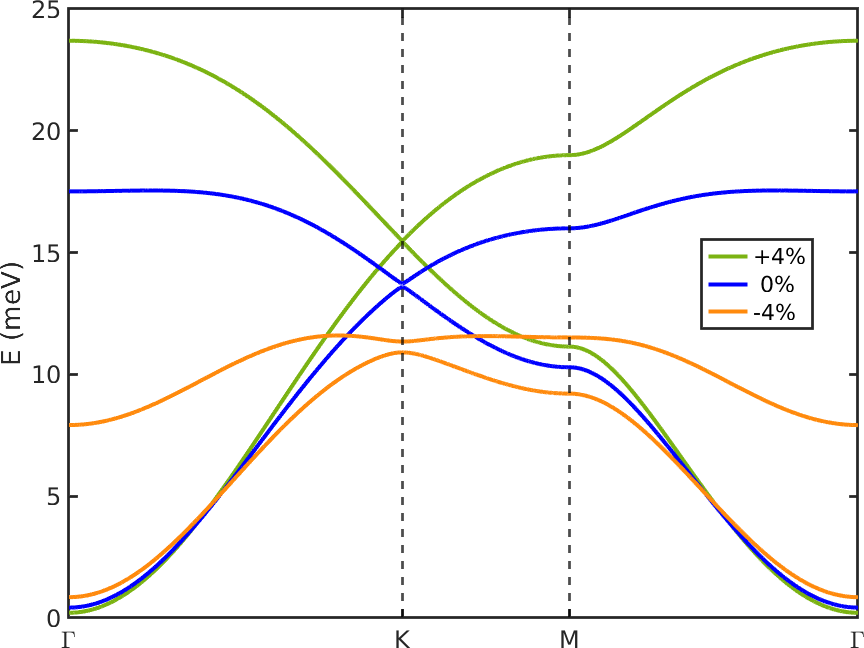}}
\hfill
\subfloat[\raggedright(b)]{\includegraphics[width=.47\textwidth]{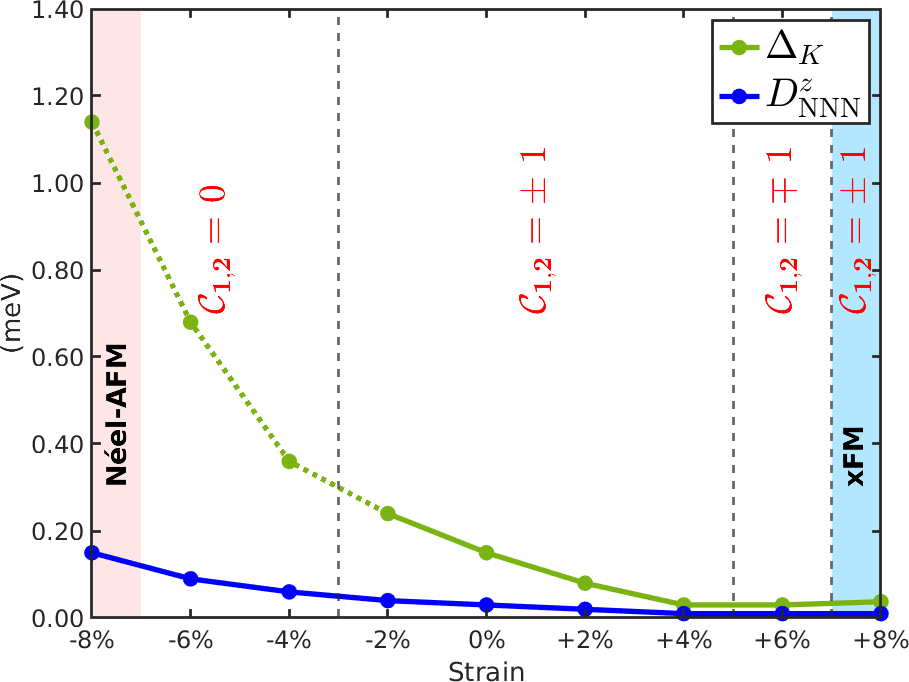}}
\hspace*{\fill}
\caption{\label{fig:biaxial_strain_dispersion} Effect of biaxial strain on the magnonic properties of monolayer CrI$_3$. (a) Shape of the magnonic dispersion for different values of applied biaxial strain. (b) Size of the magnonic bandgap (at the K point) and the out-of-plane component of the NNN DMI as a function of the applied strain. The green dotted line denotes that, although $\Delta_K$ is non-zero, there is no effective bandgap due to the lowering of the branches at the $\Gamma$ point. Corresponding Chern numbers are indicated for each topological phase, topological phase transitions are marked with black dotted lines.}
\end{figure*}

Figs.~\ref{fig:biaxial_strain_parameters}(b) and \ref{fig:biaxial_strain_parameters}(c) depict the dependence of the exchange and SIA parameters (in the Cartesian basis) on the biaxial strain. In Fig.~\ref{fig:biaxial_strain_parameters}(b), we defined the in-plane and out-of-plane exchange constants as $J_{\parallel} = [\langle \mathcal{J}_{ij}^{xx} \rangle + \langle \mathcal{J}_{ij}^{yy} \rangle]/2$ with $\langle \mathcal{J}_{ij}^{xx} \rangle \approx \langle \mathcal{J}_{ij}^{yy} \rangle$, and $J^{zz} = \langle \mathcal{J}_{ij}^{zz} \rangle$, where $\langle\rangle$ corresponds to averages taken over all appropriate $(i-j)$ pairs for the NN and NNN parameters respectively. Similarly, we define the out-of-plane exchange anisotropies portrayed in Fig.~\ref{fig:biaxial_strain_parameters}(c) as $\Delta = \langle \mathcal{J}_{ij}^{zz} \rangle - J_{\parallel}$. The dotted lines in Figs. \ref{fig:biaxial_strain_parameters}(b) and \ref{fig:biaxial_strain_parameters}(c) mark magnetic phase transitions between the spin configurations depicted in Fig.~\ref{fig:biaxial_strain_parameters}(a). The NN exchange constants decrease for applied tensile strains, while the SIA increases, eventually resulting in a transition to an in-plane FM state (xFM), marked with blue on Fig. \ref{fig:biaxial_strain_parameters}. The transition occurs at the point where $\mathcal{A}_{ii}^{zz} + \Delta_{\mathrm{NNN}} > |\Delta_{\mathrm{NN}}|$, assuring that the in-plane state is now lower in energy than the out-of-plane (zFM) one. In contrast, the exchange constants increase rapidly for compressive strains, eventually even turning positive, which leads to a transition to an in-plane N\'eel-AFM state, marked with red on Fig. \ref{fig:biaxial_strain_parameters}. Note that in the xFM state the spins are oriented along the armchair direction, while in the N\'eel-AFM state they orient themselves along the zigzag direction. The emerging magnetic phase transitions are in qualitatively good agreement with earlier reported results \cite{zheng2018}, strengthening our confidence in the obtained magnetic parameters. 

Naturally, the changes in the magnetic parameters as a consequence of the applied strain will also influence the magnonic dispersion. Fig.~\ref{fig:biaxial_strain_dispersion}(a) shows the influence of a 4\% tensile strain and a -4\% compressive strain on the shape of the dispersion. The bandwidth of the magnon spectrum will increase or decrease proportionally to the strength of the exchange interaction [i.e. increase for tensile strain, decrease for compressive strain; cf. Fig.~\ref{fig:biaxial_strain_parameters}(a)]. The size of the Goldstone gap will decrease for applied tensile strains as it is proportional to the out-of-plane magnetic anisotropy of the system [cf. Fig.~\ref{fig:biaxial_strain_parameters}(b)]. Fig.~\ref{fig:biaxial_strain_dispersion}(b) depicts the size of the magnonic bandgap (at the K point) as a function of the applied strain. It is clear that the size of $\Delta_K$ evolves proportionally to the the out-of-plane DMI component for both compressive and tensile strain. However, despite the fact that $\Delta_K$ is non-zero for all considered strains, there will be no effective bandgap for applied strains of -4\% and lower, due to the lowering of the branches at the $\Gamma$ point as can be seen from the orange curve in Fig.~\ref{fig:biaxial_strain_dispersion}(a). Applying biaxial strain not only influences the size of the bandgap but also its topology. Fig.~\ref{fig:biaxial_strain_dispersion}(b) shows the Chern numbers for the different topological phases that can be reached by applying biaxial strain. When there is no effective bandgap, the Chern numbers, of course, vanish and the system finds itself in a topologically trivial phase. For a tensile strain of about 6\% the sign of the Chern number of each band changes sign, signifying that the propagation direction of the magnonic edge states, and the corresponding thermal Hall current will reverse. Note that such a sign change is expected for decreasing DMI, which can be seen from the topological phase diagram as a function of the the DMI and the Kitaev interaction reported in Ref.~\cite{soenen2023}. When the magnetic phase changes to the xFM configuration, the Chern numbers again change sign, since now the bandgap is determined by the in-plane DMI component instead of the out-of-plane one.

\begin{figure*}[tp!]
\centering
\hspace*{\fill}
\subfloat[\raggedright(a)]{\includegraphics[width=.45\textwidth]{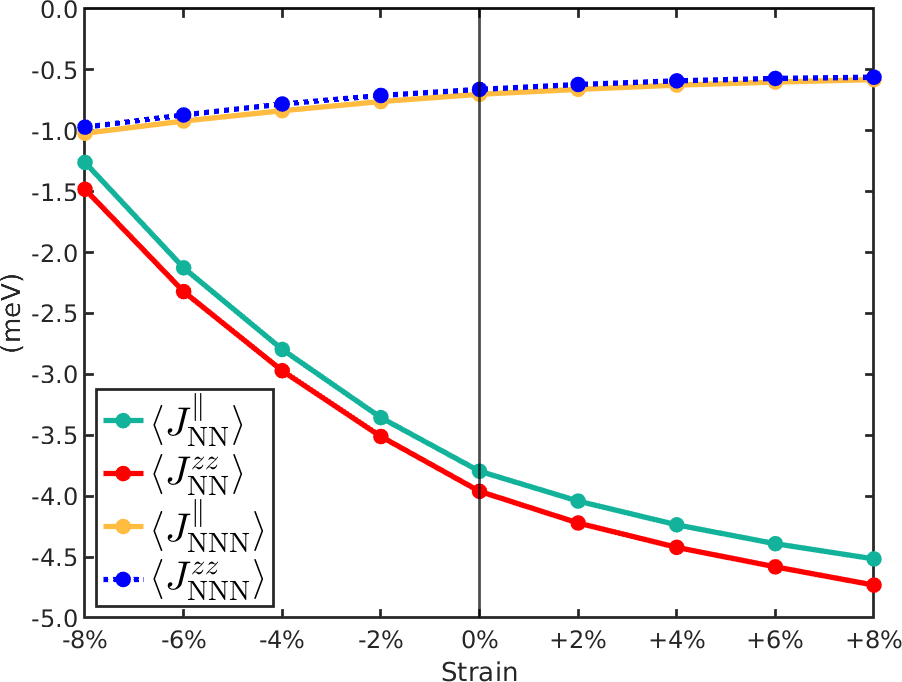}}
\hfill
\subfloat[\raggedright(b)]{\includegraphics[width=.495\textwidth]{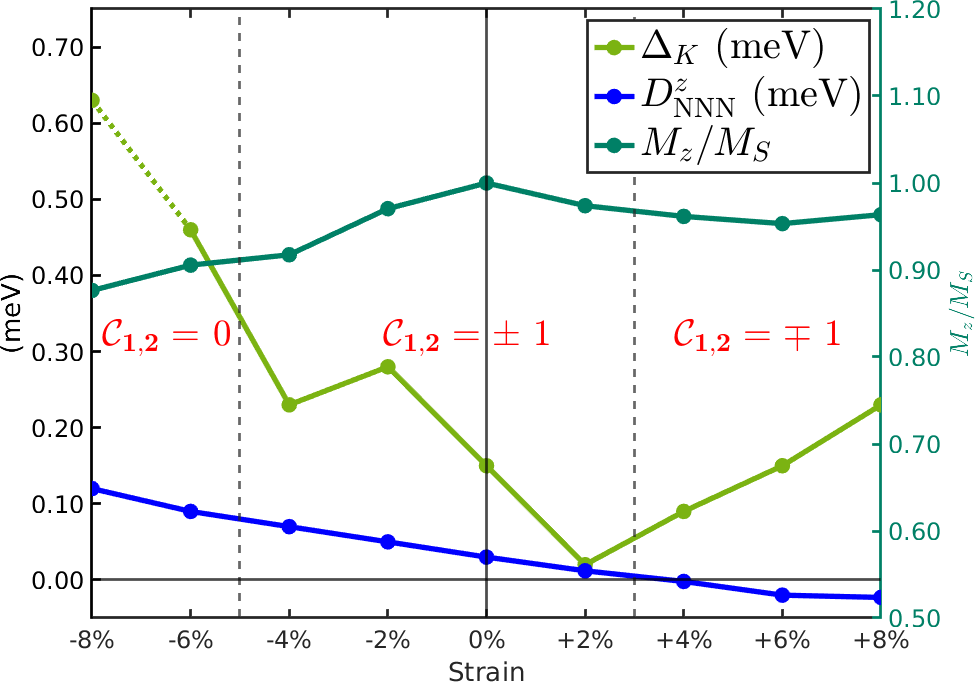}}
\hspace*{\fill}
\caption{\label{fig:uniaxial_strain_dispersion} Effect of uniaxial strain, applied along the zigzag direction of the chromium honeycomb lattice, on the magnetic and magnonic properties of monolayer CrI$_3$. (a) Evolution of the in-plane and out-of-plane exchange constants as a function of the applied strain. (b) Plot of the size of the magnonic bandgap and the out-of-plane component of the NNN DMI on the left axis, and the out-of-plane component of the magnetization on the right axis, all as a function of the applied strain. Corresponding Chern numbers are indicated for each phase, and phase transitions are marked with black dotted lines.}
\end{figure*}

\subsection{Effect of uniaxial strain}\label{sec:uniax_strain}
By applying biaxial strain, one changes the distance between the spins in monolayer CrI$_3$ while conserving its lattice symmetries. However, if we apply uniaxial strain, the rotational symmetry of the lattice will be broken, which is reflected in the resulting exchange parameters [see Fig.~\ref{fig:uniaxial_strain_dispersion}(a)]. In what follows, the uniaxial strain is applied along the zigzag direction of the chromium honeycomb lattice \footnote{Note that the atomic positions were relaxed in the direction perpendicular to the applied strain.}. For the NN exchange parameters associated with the pair oriented perpendicular to the direction of the strain, there are only minor changes in the parameter values, while the changes for the other two NN exchange parameters are more significant \cite{sup_mat}. Consequently, the overall in-plane and out-of-plane exchange constants will evolve in a similar fashion as the parameters for the biaxial strain case, but they will grow and decline at a slower rate. Another effect of the broken rotational symmetry is that the SIA is no longer quantified by only one parameter but by the full SIA matrix, which causes the easy axis of the spins to tilt with respect to the out-of-plane axis \cite{sup_mat}. Together with the non-uniform changes to the exchange parameters of the different pairs, this causes spin canting along the direction of the applied strain, i.e. the zigzag direction. 

Fig.~\ref{fig:uniaxial_strain_dispersion}(b) plots the magnonic bandgap and corresponding Chern numbers of monolayer CrI$_3$ as a function of the applied uniaxial strain. We identify the NNN DMI and the canting of the spins caused by a change in the magnetic anisotropy, the latter quantified by the out-of-plane component of the magnetization, as the two largest factors that determine the size of the bandgap. 

\begin{figure*}[tp!]
\centering
\hspace*{\fill}
\subfloat[\raggedright(a)]{\includegraphics[width=.45\textwidth]{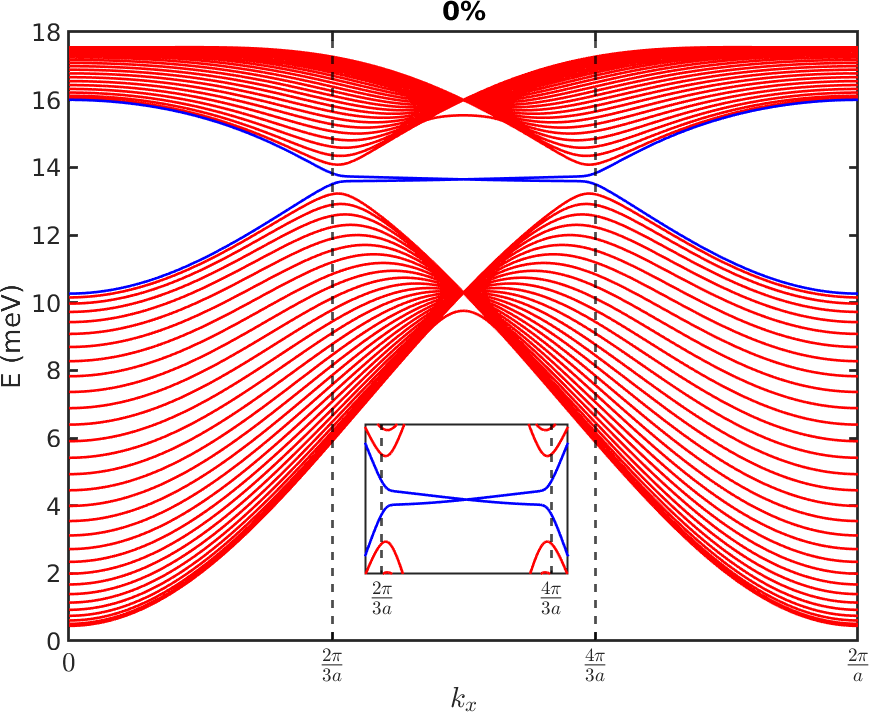}}
\hfill
\subfloat[\raggedright(b)]{\includegraphics[width=.45\textwidth]{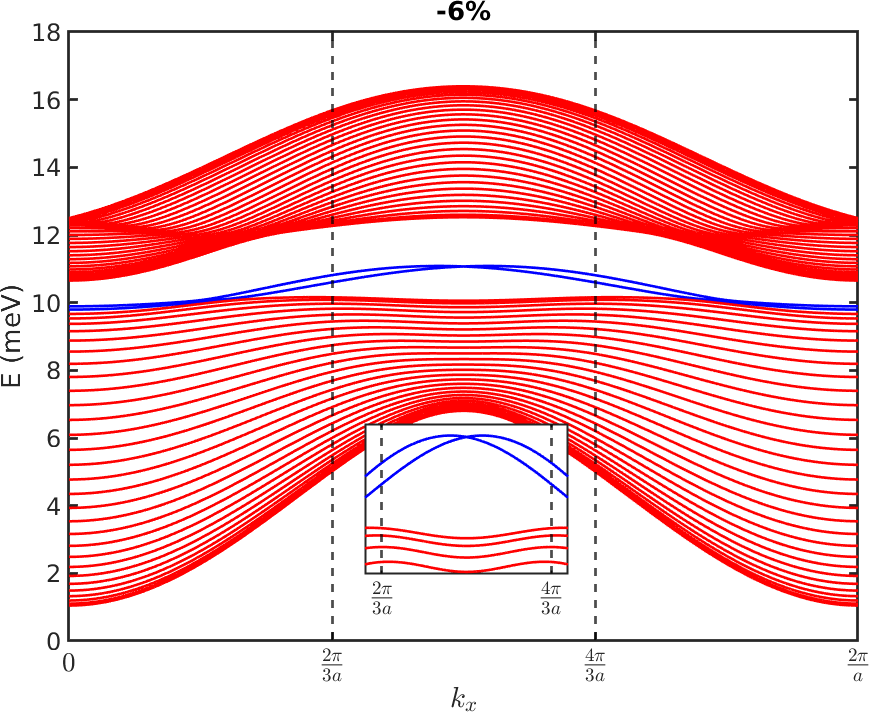}}
\hspace*{\fill}
\caption{\label{fig:edge-states} Effect of uniaxial strain on the magnonic dispersion of a quasi one-dimensional CrI$_3$ stripe (of 30 unit cells width and with zigzag edges; panel (a) shows pristine case, panel (b) the -6\% compressed stripe in the longitudinal direction). Bulk magnonic bands are depicted in red, edge states in blue.}
\end{figure*}

For compressive uniaxial strain, the bandgap has the tendency to increase proportional to the DMI. However, the spin canting has the opposite effect and suppresses the bandgap, which explains the dip in the bandgap around -4\% applied strain, where we see a similar decrease of the magnetization curve. Nevertheless, we generally notice that compressive uniaxial strain leads to an increase of the bandgap with respect to the pristine case. Remarkably, we find that for a compressive strain of -6\% and -8\% the Chern numbers vanish, despite a non-zero $\Delta_K$. To illustrate why, we investigate the edge states in a quasi one-dimensional CrI$_3$ stripe that is periodic along the $x$-direction, and has a finite width of 30 unit cells in the $y$-direction, with zigzag edges [see Fig.~\ref{fig:edge-states}]. For the pristine case, depicted in Fig.~\ref{fig:edge-states}(a), we observe two non-trivial edge modes that connect the upper and lower bands and cross at the Dirac point. However, for a compressive uniaxial strain of -6\% [Fig.~\ref{fig:edge-states}(b)], we see that the edge modes become trivial as they no longer connect the upper and lower bulk bands, hence, the Chern number equals zero. For higher compressive strain, e.g. -8\%, the effective bandgap is zeroed due to the lowering of the branches around the $\Gamma$ point, similarily to the biaxial strain case.

For tensile strain, we find that the size of the bandgap scales almost linearly with the absolute value of the DMI, while the sign change of the DMI also causes the signs of the Chern numbers to flip, again signifying a change in the propagation direction of the magnonic edge currents.

\subsection{Effect of electric gating}\label{sec:gating}
\begin{figure}[b!]
\centering
\includegraphics[width=\linewidth]{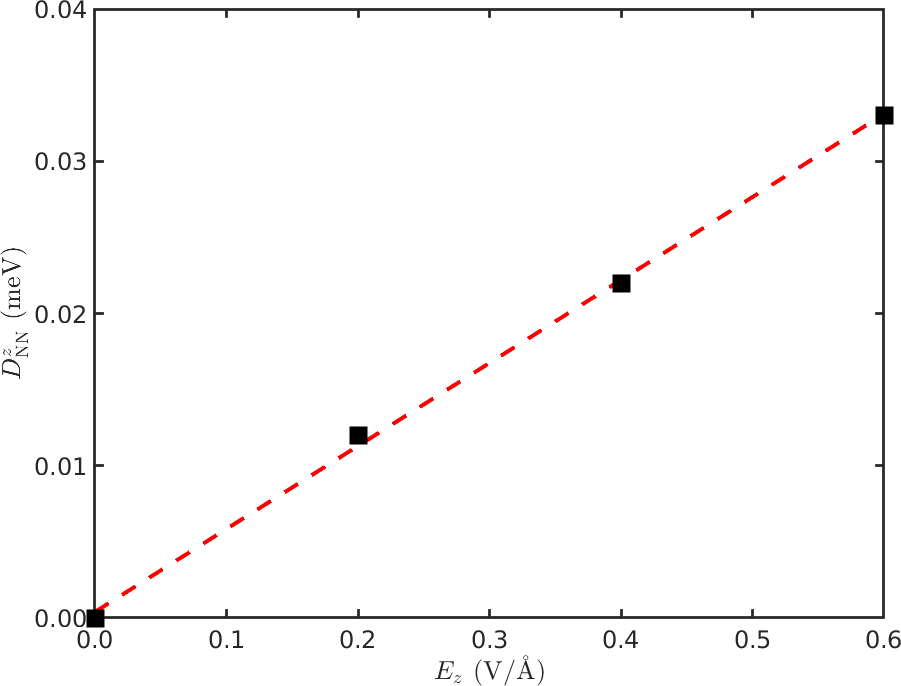}
\caption{\label{fig:efield} NN out-of-plane DMI value as a function of the applied out-of-plane electric field.}
\end{figure}

2D materials are particularly prone to vertical gating, where low applied voltages can lead to very large electric field and cause radical changes in electronic and magnetic properties. By applying such an out-of-plane electric field, we break the inversion symmetry of monolayer CrI$_3$, allowing for the emergence of DMI even between the NN spins. However, as shown in Fig.~\ref{fig:efield}, the resulting DMI is rather small even at relatively high applied fields. At the same time, we found that all other magnetic parameters remain practically unaffected by the applied field (details given in Supplementary Material \cite{sup_mat}). Our calculations show that the NN DMI scales linearly with the applied electric field, reaching values of the same order of magnitude as the ones reported in Ref.~\cite{ghosh2020}. The found typical magnitude of these DMI values is far insufficient to instigate any non-collinear spin textures in CrI$_3$. Furthermore, the found small increase in NN DMI had weak impact on the spin wave dispersion. Due to the electric field, the magnons pick up a geometric phase which will shift the Dirac surface states up or down depending on the polarity of the field with respect to the magnetization. This phenomenon is called the AC effect, and it has been proposed as a possible source of evidence to confirm the topological properties of CrI$_3$ in experiment \cite{kondo2021}. However, according to our calculations, these shifts, which are below 0.04 meV even for a large electric field of $E_z = 0.6$ V/\AA{}, are too small to be observed experimentally.    

\subsection{Effect of magnetic field}\label{sec:mag_field}
In preceding sections, we showed that applying strain or an electric field to monolayer CrI$_3$ tunes its exchange parameters and, consequently, its magnonic dispersion. In this section, we demonstrate that by applying an external magnetic field, which merely changes the spin orientation and leaves the exchange parameters intact, one can also manipulate the spin-wave behavior. By investigating the hysteresis curve of monolayer CrI$_3$ in an out-of-plane magnetic field, we first show that reversal of the magnetization results in a sign change of the Chern numbers. Furthermore, we demonstrate that an in-plane magnetic field can lead to a similar sign change, or can even fully close the magnonic bandgap, depending on the orientation of the applied field with respect to the crystallographic directions of the lattice. 

\begin{figure}[tp!]
\centering
\includegraphics[width=0.9\linewidth]{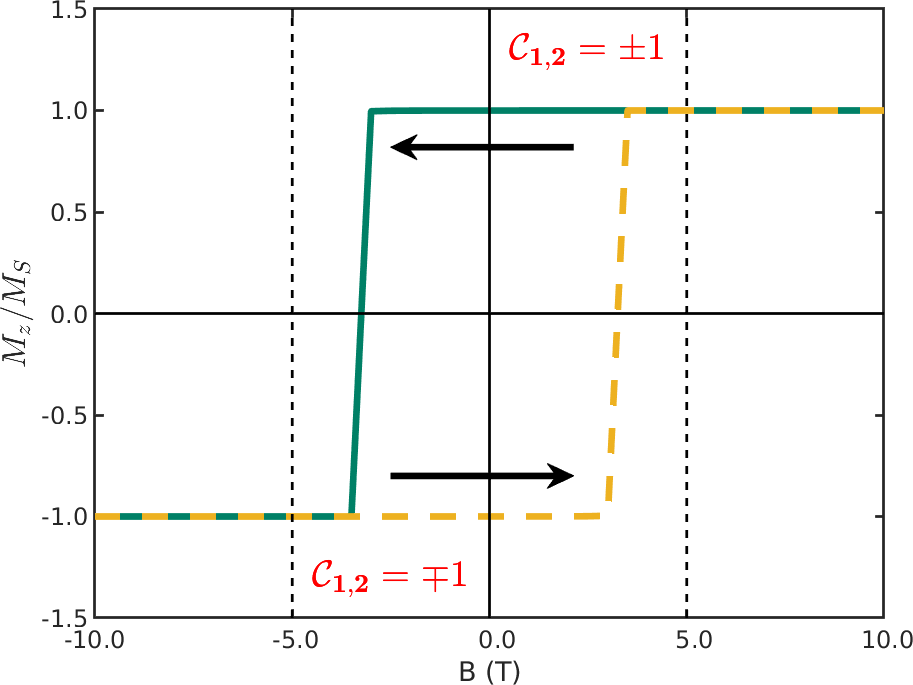}
\caption{\label{fig:hysteresis} Hysteresis curve of monolayer CrI$_3$ under applied out-of-plane magnetic field. Corresponding Chern numbers are indicated for both magnetization orientations.}
\end{figure}

\emph{Out-of-plane field} --- Applying a uniform out-of-plane magnetic field has no effect on the shape of the magnonic dispersion, however, the Goldstone gap will increase (decrease) for magnetic fields that have a parallel (opposite) polarity with respect to the magnetization direction \cite{soenen2023}. Fig.~\ref{fig:hysteresis} shows the hysteresis curve of monolayer CrI$_3$, predicting that beyond the coercive field of 3.25 T \footnote{One should note that the coercive field values reported here are significantly larger than the ones observed in experiments \cite{huang2017}, which we attribute to the idealized simulations conditions (e.g. T = 0 K, infinitely periodic yet defect-less lattice, etc.), which are unrealistic in practice.} the Goldstone gap closes, causing the magnetization to flip. Notice that a flip of the magnetization causes a sign change of the Chern numbers, reversing the propagation direction of the magnonic edge states and the resulting heat current. 

\begin{figure}[tp!]
\centering
\includegraphics[width=0.9\linewidth]{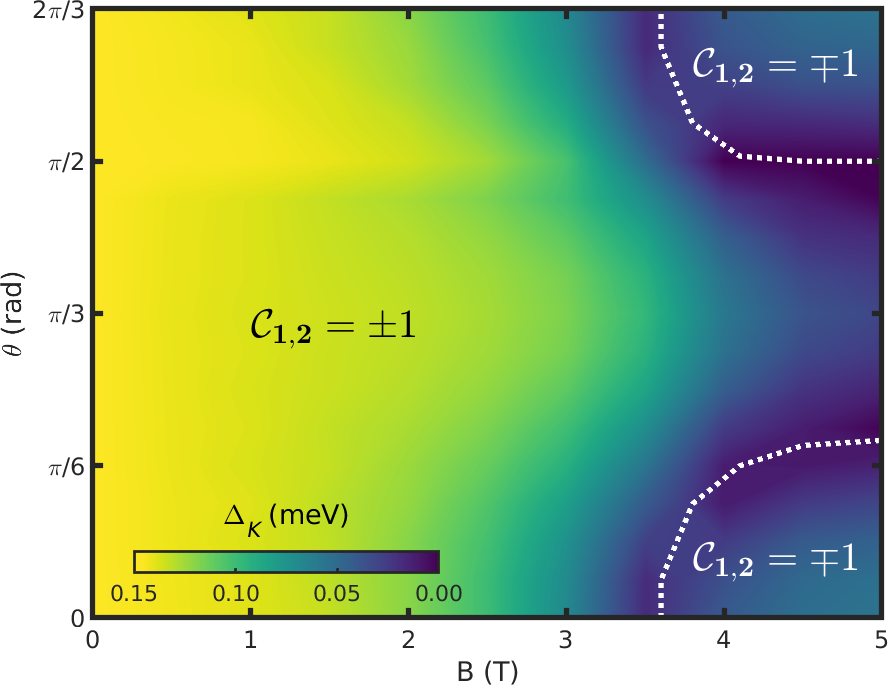}
\caption{\label{fig:mag_field} Size of the magnonic bandgap $\Delta_K$ as a function of the magnitude and direction of an applied in-plane magnetic field. Corresponding Chern numbers and the topological phase boundaries are indicated.}
\end{figure}

\emph{In-plane field} --- Fig.~\ref{fig:mag_field} shows the effect of a uniform in-plane magnetic field on monolayer CrI$_3$, by plotting the size of the magnonic bandgap as a function of the magnitude of the field $B$ and the azimuthal angle $\theta$ between the armchair direction of the lattice and the direction of the field. As the strength of field is increased, the spins tilt towards the direction of the applied field. We find that the saturation field, for which all the spins attain in-plane polarization, lies in the range 3.0-3.5 T, depending on the direction of the applied field, which is in good agreement with experimentally reported values of approximately 3 T for bulk CrI$_3$ \cite{mcguire2015,chen2021}. Notice that the size of the magnonic bandgap decreases proportionally to the out-of-plane polarization of the spins. For strong fields, when the spins are fully polarized in the lattice plane, the bandgap can even close entirely, which has also been shown theoretically in the Heisenberg-Kitaev model \cite{chen2021}. Note, however, that the effect of the field on the bandgap depends on the direction in which it is applied. For fields applied along or under small angle $\theta$ to the armchair direction, we find a parametric $(B,\theta)$ region, marked with a white dotted line on Fig.~\ref{fig:mag_field}, where the bandgap does not fully close. Within this region, the bandgap is small but non-zero, accompanied by a sign change of the Chern numbers. These results seem to be in qualitatively good agreement with Ref.~\cite{zhang2021}, who reported a similar sign change of the Chern numbers and the thermal Hall conductivity as a function of an in-plane magnetic field for arbitrarily varied parameters in a general spin model for 2D honeycomb ferromagnets (whereas here we obtain the material-specific microscopic exchange parameters from first-principles calculations). Finally, notice that due to the breaking of in-plane honeycomb symmetry by the applied field, the bandgap does not occur exactly at the $K$ point, but is slightly shifted instead, as seen in earlier studies on bilayer and bulk CrI$_3$ \cite{soenen2023, gorni-arxiv}.

\section{Conclusions}\label{sec:conclu}
Using a Heisenberg spin model parametrized from first principles, we characterized the magnonic dispersion of monolayer CrI$_3$ under various external stimuli accessible in experiment, showing that both the size and topology of the magnonic bandgap can thereby be tuned. By applying biaxial strain, uniaxial strain, or an out-of-plane or in-plane magnetic field, we demonstrated tunability of the magnon topology between multiple phases, either switching between topologically trivial and topologically non-trivial phases, or reversing the sign of the Chern numbers between two topologically non-trivial phases. In topological magnonic devices, these findings could be employed to switch on/off or reverse the magnon edge current or the thermal Hall current, or simply tune the size of the bandgap and manipulate the bandwidth in which the device operates. 

Counterintuitively, we found that electric gating exerts a negligible effect on the magnetic properties and magnonic band structure of monolayer CrI$_3$. This result is rather disappointing with respect to electronic control of related magnonic devices and requires experimental validation.  

Although, several neutron scattering measurements of 2D materials confirmed the presence of magnonic bandgaps that are predicted to posses topological properties \cite{chen2018,chen2021,cai2021,zhu2021,chisnell2015}, the thermal magnon Hall effect remained undetected to date, which would have been a decisive characteristic signature for the existence of topological magnons in 2D materials. This observational hiatus is likely due to the condensation of magnons in the bottom of the bands as they follow Bose-Einstein statistics, however, recent work suggests that the amplification of the edge states using very specifically tailored electromagnetic fields could provide a route towards experimental verification of the thermal Hall effect \cite{malz2019}, which may then further be employed to validate our results in CrI$_3$. 

On the theory front, there are also several open challenges requiring further research. An important unsolved problem is achieving more accurate predictions of the size of the magnonic bandgap. In this regard, efforts should be directed towards understanding and modelling of the magnon-phonon coupling. Another research direction should be to advance the study of topological magnonics in materials with different lattice types or spin configurations, bearing in mind the pertinent interest in using periodic spin textures as magnonic crystals \cite{ma2015,yu2021}.   
 
\begin{acknowledgments}
We thank Cihan Bacaksiz, Ali Ghojavand, Denis \v{S}abani, and Cem Sevik for useful discussions. This work was supported by the Research Foundation-Flanders (FWO-Vlaanderen, Grant No. 11O1423N) and the Special Research Funds of the University of Antwerp (BOF-UA). The computational resources used in this work were provided by the VSC (Flemish Supercomputer Center), funded by Research Foundation-Flanders (FWO) and the Flemish Government -- department EWI.
\end{acknowledgments}

\begin{figure}[tp!]
\centering
\subfloat[(a) CrBr$_3$]{\includegraphics[width=.9\linewidth]{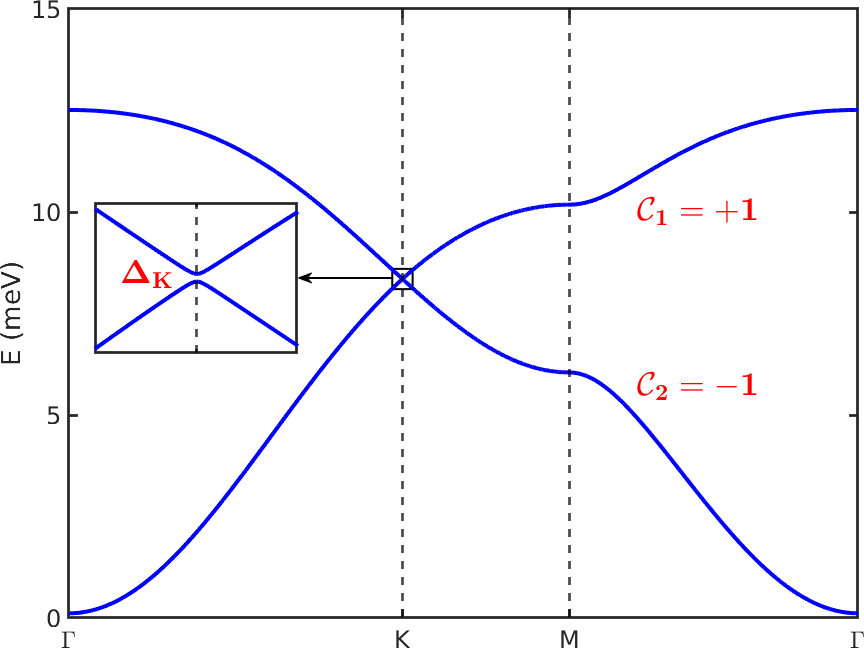}} \\
\subfloat[(b) CrCl$_3$]{\includegraphics[width=.9\linewidth]{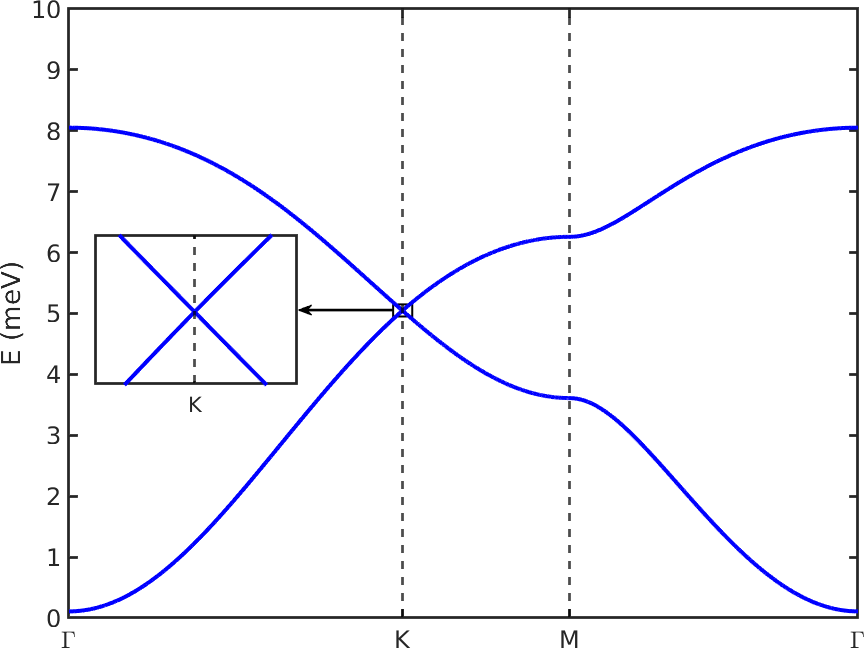}}
\caption{\label{fig:crx3} Magnonic dispersion of monolayers CrBr$_3$ (a) and CrCl$_3$ (b). For CrBr$_3$, a very small bandgap of $\Delta_K$ = 0.03 meV is present at the $K$ point, and corresponding Chern numbers are indicated for each band. CrCl$_3$ exhibits a Dirac cone at the $K$ point and, thus, a trivial topology.}
\end{figure}

\appendix
\section{DFT calculations}\label{app:dft}
The Vienna \emph{ab initio} simulation package (VASP) \cite{vasp1,vasp2,vasp3} is used to conduct DFT calculations, employing the generalized gradient approximation (GGA) by implementing the Perdew-Burke-Ernzerhof (PBE) \cite{pbe} exchange-correlation functional, and the projector augmented wave (PAW) \cite{paw} method. To incorporate a vdW correction term, we utilize the D2 method of Grimme \cite{grimme}. The SOC is included in all calculations \cite{soc}. To improve the description of the strongly-correlated d-electrons, we implement the GGA+U method in the rotationally invariant form \cite{dudarev} by adding an on-site Coulomb interaction of U$^{\mathrm{eff}}$ = U - J = 2.8037 eV to the d-orbitals of the chromium atoms. The latter value was calculated using a linear response method. During structural relaxations, we use a plane-wave energy cutoff of 700 eV, during other 4SM calculations we can safely lower the cutoff to 300 eV \footnote{In an earlier study on monolayer CrI$_3$ \cite{soenen2023}, we performed convergence tests which demonstrate that we can lower the energy cutoff to 300 eV without qualitative impact and very limited quantitative impact on the obtained exchange parameters.}. Due to the periodic boundary conditions of VASP, we implement a unit cell length of c = 20 \AA\ in the out-of-plane direction to include enough vacuum between periodic images. In the 4SM calculations, we require a 3 $\times$ 3 $\times$ 1 supercell to assure fully converged magnetic parameters. Smaller supercells result in an overestimation of the exchange constants due to artificial interactions between periodic images. For the BZ integration, we use a Gaussian smearing of 0.01 eV. During structural relaxation, a k-point sampling grid of $15 \times 15 \times 1$ is utilized, while the use of supercells in the 4SM calculations permits the use of a smaller 3 $\times$ 3 $\times$ 1 grid to minimize the computational cost. To investigate the dynamical stability of strained CrI$_3$, we calculated the phonon dispersion for different applied strains. The obtained phonon band structures showed no sign of instabilities for the strain values considered in this work.

\section{Magnonic dispersion of monolayer CrBr$_\mathbf{3}$ and CrCl$_\mathbf{3}$}\label{app:crx3}
In order to investigate the effect of the SOC on the magnon topology, and in search of realizing Dirac magnons in a magnetic monolayer, we detail the spin-wave dispersion of CrBr$_3$ and CrCl$_3$.

The magnonic bandgap in CrI$_3$ originates from the DMI and the Kitaev interaction, which are both a consequence of the strong SOC due to its heavy iodine ligands. Other chromium trihalide monolayers, i.e. CrBr$_3$ or CrCl$_3$, have lighter ligands and, thus, a smaller SOC, leading to heavily reduced magnetic parameters of respectively, $D^z_\mathrm{NNN} = 0.00$ meV and $K_\mathrm{NN} = 0.15$ meV for CrBr$_3$, and $D^z_\mathrm{NNN} = 0.00$ meV and $K_\mathrm{NN} = 0.03$ meV for CrCl$_3$. A full summary of all the magnetic parameters for CrBr$_3$ and CrCl$_3$ can be found in the Supplementary Material \cite{sup_mat}. The obtained magnetic parameters lead to an out-of-plane FM ground state for both CrBr$_3$ and CrCl$_3$, although the latter is in disagreement with experimental work on bulk CrCl$_3$ \cite{mcguire2017}, our work is consistent with earlier DFT-based studies of monolayer CrCl$_3$ \cite{dupont2021}. The spin wave dispersion for both materials is depicted in Figs. \ref{fig:crx3}(a) and \ref{fig:crx3}(b) for CrBr$_3$ and CrCl$_3$ respectively. The dispersion has a similar shape to the one of CrI$_3$, however, the reduced DMI and Kitaev values now result in a tiny topological bandgap of $\Delta_K$ = 0.03 meV for CrBr$_3$ and a Dirac cone for CrCl$_3$. Since the DMI is zero is both materials, we attribute the origin of the bandgap in CrBr$_3$ solely to the Kitaev interaction. The Chern numbers for CrBr$_3$ are indicated in the figure. Our calculations for CrCl$_3$ confirm that the gapless Dirac magnons recently observed in the bulk \cite{chen2022,schneeloch2022}, persist down to the monolayer limit.

\end{document}


\title{Tunable magnon topology in monolayer CrI$_\mathbf{3}$ under external stimuli\\ SUPPLEMENTAL MATERIAL}

\author{Maarten Soenen}
\affiliation{Department of Physics \& NANOlab Center of Excellence, University of Antwerp, Groenenborgerlaan 171, B-2020 Antwerp, Belgium}

\author{Milorad V. Milo\v{s}evi\'c}
\email{milorad.milosevic@uantwerpen.be}
\affiliation{Department of Physics \& NANOlab Center of Excellence, University of Antwerp, Groenenborgerlaan 171, B-2020 Antwerp, Belgium}
\affiliation{Instituto de Física, Universidade Federal de Mato Grosso, Cuiabá, Mato Grosso 78060-900, Brazil}

\date{\today}

\maketitle

\tableofcontents

\begin{figure}[b!]
\includegraphics[width=0.90\linewidth]{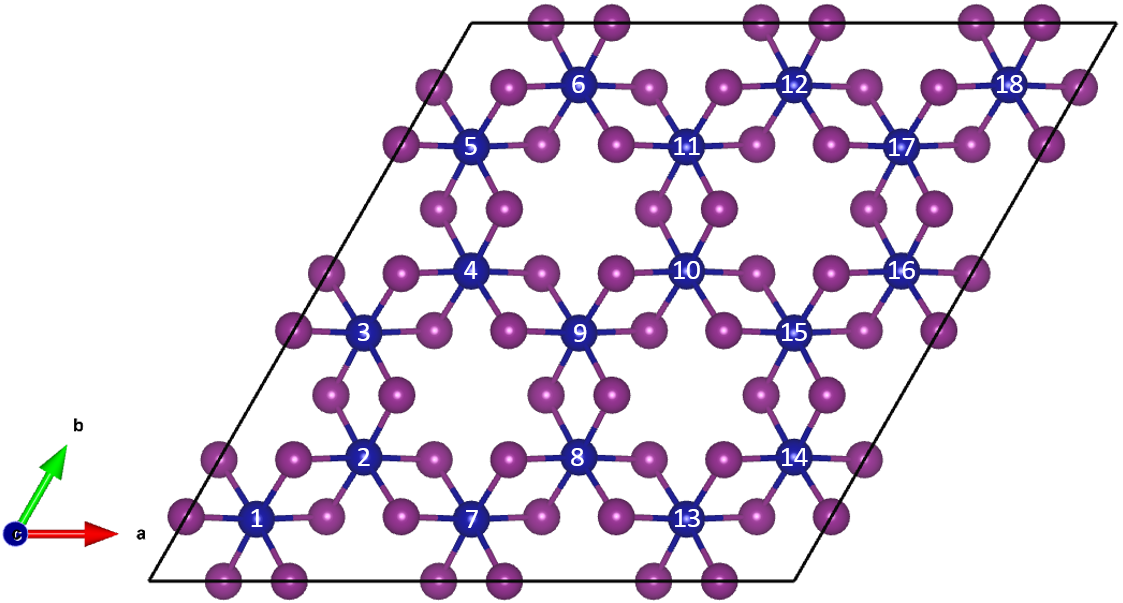}
\caption{\label{fig:pairs} Top view of the crystal structure of monolayer CrI$_3$. The chromium and iodine atoms are depicted with dark blue and purple spheres respectively. Each spin site is labeled with a number. The unit cell is marked with a solid black line. The crystal structure was drawn using \textsc{vesta} \cite{VESTA}.}
\end{figure}

\newpage
\section{Parameters for the pristine chromium trihalide monolayers}
\begin{table}[!ht]
\caption{\label{tab:crx3-ex} Calculated exchange parameters for pristine CrI$_3$, CrBr$_3$, and CrCl$_3$ monolayers. Corresponding (i-j) pairs are indicated in Fig. \ref{fig:pairs}. All values are given in a Cartesian basis.}
\begin{tabular*}{\linewidth}{@{\extracolsep{\fill}} ccccccccccc }
\hline\hline
& Pair & $\mathcal{J}^{xx}_{ij}$ & $\mathcal{J}^{yy}_{ij}$ & $\mathcal{J}^{zz}_{ij}$ & $\mathcal{J}^{xy}_{ij}$ & $\mathcal{J}^{yx}_{ij}$ & $\mathcal{J}^{xz}_{ij}$ & $\mathcal{J}^{zx}_{ij}$ & $\mathcal{J}^{yz}_{ij}$ & $\mathcal{J}^{zy}_{ij}$ \\
& (i-j)   & (meV) & (meV) & (meV) & (meV) & (meV) & (meV) & (meV) & (meV) & (meV) \\ \hline
CrI$_3$ &         & & & & & & & & & \\
& (1-2) & -4.34 & -3.24 & -3.96 & 0.00 & 0.00 & 0.00 & 0.00 & -0.65 & -0.65 \\ 
& (2-3) & -3.53 & -4.07 & -3.97 & -0.50 & -0.50 & -0.57 & -0.57 & 0.33 & 0.33 \\
& (2-7) & -3.53 & -4.07 & -3.97 & 0.50 & 0.50 & 0.57 & 0.57 & 0.33 & 0.33 \\
&         & & & & & & & & & \\
& (1-3) & -0.67 & -0.73 & -0.66 & 0.08 &  0.03 & 0.09 & 0.03 & -0.01 & 0.09 \\
& (1-7) & -0.67 & -0.73 & -0.66 & -0.08 & -0.03 & -0.09 & -0.03 & -0.01 & 0.09 \\
& (9-13) & -0.76 & -0.64 & -0.66 & 0.03 & -0.03 & -0.05 & 0.05 & -0.08 & -0.08 \\ \hline
CrBr$_3$ &         & & & & & & & & & \\
& (1-2)   & -2.81 & -2.70 & -2.79 &  0.00 &  0.00 &  0.00 &  0.00 & -0.06 & -0.06 \\
& (2-3)   & -2.73 & -2.78 & -2.78 & -0.05 & -0.05 & -0.06 & -0.05 &  0.03 &  0.03 \\
& (2-7)   & -2.73 & -2.78 & -2.78 &  0.05 &  0.05 &  0.06 &  0.06 &  0.03 &  0.03 \\
&         & & & & & & & & & \\
& (1-3)   & -0.30 & -0.30 & -0.30 & 0.00 &  0.01 &  0.01 & 0.00 & -0.01 &  0.02 \\
& (1-7)   & -0.30 & -0.30 & -0.30 & 0.00 & -0.01 & -0.01 & 0.00 & -0.01 &  0.02 \\
& (9-13)  & -0.31 & -0.30 & -0.30 & 0.00 &  0.00 & -0.02 & 0.02 & -0.01 & -0.01 \\ \hline
CrCl$_3$ &         & & & & & & & & & \\
& (1-2)   & -1.76 & -1.77 & -1.82 & 0.00 & 0.00 & 0.00 & 0.00 & 0.00 & 0.00 \\
& (2-3)   & -1.77 & -1.76 & -1.77 & 0.00 & 0.00 & 0.00 & 0.00 & 0.00 & 0.00 \\
& (2-7)   & -1.77 & -1.76 & -1.77 & 0.00 & 0.00 & 0.00 & 0.00 & 0.00 & 0.00 \\
&         & & & & & & & & & \\
& (1-3)   & -0.14 & -0.14 & -0.14 & 0.00 & 0.00 & 0.00 & 0.00 & 0.00 & 0.00 \\
& (1-7)   & -0.14 & -0.15 & -0.14 & 0.00 & 0.00 & 0.00 & 0.00 & 0.00 & 0.00 \\
& (9-13)  & -0.15 & -0.14 & -0.19 & 0.00 & 0.00 & 0.00 & 0.00 & 0.00 & 0.00 \\ \hline \hline
\end{tabular*}
\end{table}

\begin{table}[!ht]
\caption{\label{tab:crx3-sia} Calculated SIA parameter for pristine CrI$_3$, CrBr$_3$, and CrCl$_3$ monolayers. Values are given in meV and are represented in a Cartesian basis.}
\begin{tabular}{lr}
\hline\hline
        & \qquad $\mathcal{A}^{zz}_{ii}$  \\ \hline
CrI$_3$  & -0.08 \\
CrBr$_3$ & -0.02 \\
CrCl$_3$ & 0.00 \\ \hline \hline
\end{tabular}
\end{table}
\clearpage

\newpage
\section{Parameters for monolayer CrI$_\mathbf{3}$ under biaxial strain}
\begin{table}[!ht]
\caption{\label{tab:biaxial_strain_nn} Calculated exchange parameters for the NN pairs of monolayer CrI$_3$ under biaxial strain. Corresponding (i-j) pairs are indicated in Fig. \ref{fig:pairs}. All values are given in a Cartesian basis.}
\begin{tabular*}{\linewidth}{@{\extracolsep{\fill}} crccccccccc }
\hline\hline
Pair & Strain & $\mathcal{J}^{xx}_{ij}$ & $\mathcal{J}^{yy}_{ij}$ & $\mathcal{J}^{zz}_{ij}$ & $ \mathcal{J}^{xy}_{ij}$ & $ \mathcal{J}^{yx}_{ij}$ & $\mathcal{J}^{xz}_{ij}$ & $\mathcal{J}^{zx}_{ij}$ & $ \mathcal{J}^{yz}_{ij}$ & $\mathcal{J}^{zy}_{ij}$ \\
(i-j)   & (\%) & (meV) & (meV) & (meV) & (meV) & (meV) & (meV) & (meV) & (meV) & (meV) \\ \hline
(1-2)   &     & \\
        & -8 &  1.52 &  2.22 &  1.35 & 0.04 & 0.00 & 0.00 & 0.02 & -0.50 & -0.50 \\
        & -6 & -0.48 &  0.34 & -0.35 & 0.00 & 0.00 & 0.00 & 0.00 & -0.54 & -0.54 \\
        & -4 & -2.04 & -1.12 & -1.78 & 0.00 & 0.00 & 0.00 & 0.00 & -0.58 & -0.58 \\
        & -2 & -3.25 & -2.26 & -2.95 & 0.00 & 0.00 & 0.00 & 0.00 & -0.61 & -0.61 \\
        &  0 & -4.34 & -3.24 & -3.96 & 0.00 & 0.00 & 0.00 & 0.00 & -0.65 & -0.65 \\
        & +2 & -5.13 & -3.93 & -4.71 & 0.00 & 0.00 & 0.00 & 0.00 & -0.69 & -0.69 \\
        & +4 & -5.88 & -4.53 & -5.39 & 0.00 & 0.00 & 0.00 & 0.00 & -0.74 & -0.74 \\
        & +6 & -6.56 & -5.06 & -6.01 & 0.00 & 0.00 & 0.00 & 0.00 & -0.80 & -0.80 \\
        & +8 & -7.20 & -5.52 & -6.58 & 0.00 & 0.00 & 0.00 & 0.00 & -0.86 & -0.86 \\ \hline
(2-3)        &     & & & & & & & & & \\
        & -8 &  2.01 &  1.76 &  1.64 & -0.37 & -0.37 & -0.39 & -0.39 & 0.20 & 0.21 \\
        & -6 &  0.14 & -0.21 & -0.36 & -0.40 & -0.40 & -0.46 & -0.46 & 0.25 & 0.25 \\
        & -4 & -1.34 & -1.79 & -1.80 & -0.39 & -0.39 & -0.50 & -0.50 & 0.28 & 0.28 \\
        & -2 & -2.52 & -3.00 & -2.95 & -0.44 & -0.44 & -0.53 & -0.53 & 0.30 & 0.30 \\
        & 0  & -3.53 & -4.07 & -3.97 & -0.50 & -0.50 & -0.57 & -0.57 & 0.33 & 0.33 \\
        & +2 & -4.25 & -4.84 & -4.71 & -0.54 & -0.54 & -0.59 & -0.59 & 0.34 & 0.34 \\
        & +4 & -4.89 & -5.56 & -5.39 & -0.60 & -0.60 & -0.64 & -0.64 & 0.37 & 0.37 \\
        & +6 & -5.47 & -6.20 & -6.01 & -0.67 & -0.67 & -0.69 & -0.69 & 0.40 & 0.40 \\
        & +8 & -5.97 & -6.80 & -6.57 & -0.75 & -0.75 & -0.75 & -0.75 & 0.43 & 0.43 \\ \hline
(2-7)   &     & & & & & & & & & \\
        & -8 &  2.01 &  1.76 &  1.64 & 0.37 & 0.37 & 0.39 & 0.39 & 0.20 & 0.23 \\
        & -6 &  0.14 & -0.21 & -0.36 & 0.40 & 0.40 & 0.46 & 0.46 & 0.25 & 0.25 \\
        & -4 & -1.34 & -1.79 & -1.80 & 0.39 & 0.39 & 0.50 & 0.50 & 0.28 & 0.28 \\
        & -2 & -2.52 & -3.00 & -2.95 & 0.44 & 0.44 & 0.53 & 0.53 & 0.30 & 0.30 \\
        & 0  & -3.53 & -4.07 & -3.97 & 0.50 & 0.50 & 0.57 & 0.57 & 0.33 & 0.33 \\
        & +2 & -4.25 & -4.84 & -4.71 & 0.54 & 0.54 & 0.59 & 0.59 & 0.34 & 0.34 \\
        & +4 & -4.89 & -5.56 & -5.39 & 0.60 & 0.60 & 0.64 & 0.64 & 0.37 & 0.37 \\
        & +6 & -5.47 & -6.20 & -6.01 & 0.67 & 0.67 & 0.69 & 0.69 & 0.40 & 0.40 \\
        & +8 & -5.97 & -6.80 & -6.57 & 0.75 & 0.75 & 0.75 & 0.75 & 0.43 & 0.43 \\ \hline \hline
\end{tabular*}
\end{table}
\clearpage

\newpage
\begin{table}[!t]
\caption{\label{tab:biaxial_strain_nnn} Calculated exchange parameters for the NNN pairs of monolayer CrI$_3$ under biaxial strain. Corresponding (i-j) pairs are indicated in Fig. \ref{fig:pairs}. All values are given in a Cartesian basis.}
\begin{tabular*}{\linewidth}{@{\extracolsep{\fill}} crccccccccc }
\hline\hline
Pair & Strain & $\mathcal{J}^{xx}_{ij}$ & $\mathcal{J}^{yy}_{ij}$ & $\mathcal{J}^{zz}_{ij}$ & $ \mathcal{J}^{xy}_{ij}$ & $ \mathcal{J}^{yx}_{ij}$ & $\mathcal{J}^{xz}_{ij}$ & $\mathcal{J}^{zx}_{ij}$ & $ \mathcal{J}^{yz}_{ij}$ & $\mathcal{J}^{zy}_{ij}$ \\
(i-j)   & (\%) & (meV) & (meV) & (meV) & (meV) & (meV) & (meV) & (meV) & (meV) & (meV) \\ \hline
(1-3)   &     & & & & & & & & & \\
        & -8 & -1.36 & -1.41 & -1.25 & 0.20 & -0.08 & 0.11 & -0.01 & -0.07 & 0.18 \\
        & -6 & -1.16 & -1.23 & -1.14 & 0.15 & -0.04 & 0.19 & 0.05 & -0.07 & 0.14 \\
        & -4 & -0.97 & -1.03 & -0.95 & 0.12 & -0.01 & 0.13 & 0.02 & -0.05 & 0.12 \\
        & -2 & -0.81 & -0.87 & -0.79 & 0.09 &  0.02 & 0.11 & 0.03 & -0.02 & 0.10 \\
        & 0  & -0.67 & -0.73 & -0.66 & 0.08 &  0.03 & 0.09 & 0.03 & -0.01 & 0.09 \\
        & +2 & -0.57 & -0.64 & -0.57 & 0.07 &  0.04 & 0.09 & 0.04 &  0.00 & 0.07 \\
        & +4 & -0.49 & -0.56 & -0.50 & 0.07 &  0.05 & 0.08 & 0.04 &  0.01 & 0.06 \\
        & +6 & -0.44 & -0.51 & -0.45 & 0.07 &  0.06 & 0.07 & 0.05 &  0.02 & 0.05 \\
        & +8 & -0.40 & -0.47 & -0.42 & 0.08 &  0.06 & 0.07 & 0.06 &  0.03 & 0.04 \\ \hline
(1-7)   &     & & & & & & & & & \\
        & -8 & -1.36 & -1.42 & -1.19 & -0.20 &  0.08 & -0.11 &  0.01 & -0.10 & 0.18 \\
        & -6 & -1.16 & -1.23 & -1.14 & -0.15 &  0.04 & -0.15 & -0.05 & -0.07 & 0.14 \\
        & -4 & -0.97 & -1.03 & -0.95 & -0.12 &  0.01 & -0.13 & -0.02 & -0.05 & 0.12 \\
        & -2 & -0.81 & -0.87 & -0.79 & -0.09 & -0.02 & -0.11 & -0.03 & -0.02 & 0.10 \\
        & 0  & -0.67 & -0.73 & -0.66 & -0.08 & -0.03 & -0.09 & -0.03 & -0.01 & 0.09 \\
        & +2 & -0.57 & -0.64 & -0.57 & -0.08 & -0.04 & -0.09 & -0.04 &  0.00 & 0.07 \\
        & +4 & -0.49 & -0.56 & -0.50 & -0.07 & -0.05 & -0.08 & -0.04 &  0.01 & 0.06 \\
        & +6 & -0.44 & -0.51 & -0.45 & -0.07 & -0.06 & -0.05 & -0.07 &  0.02 & 0.05 \\
        & +8 & -0.40 & -0.47 & -0.42 & -0.08 & -0.06 & -0.07 & -0.06 &  0.03 & 0.04 \\ \hline
(9-13)  &     & & & & & & & & & \\
        & -8 & -1.41 & -1.32 & -1.30 & 0.18 & -0.18 & -0.12 & 0.16 & -0.08 & -0.08 \\
        & -6 & -1.28 & -1.13 & -1.13 & 0.09 & -0.09 & -0.13 & 0.13 & -0.09 & -0.09 \\
        & -4 & -1.07 & -0.94 & -0.94 & 0.06 & -0.06 & -0.09 & 0.09 & -0.09 & -0.09 \\
        & -2 & -0.90 & -0.78 & -0.79 & 0.04 & -0.04 & -0.07 & 0.07 & -0.08 & -0.08 \\
        & 0  & -0.76 & -0.64 & -0.66 & 0.03 & -0.03 & -0.05 & 0.05 & -0.08 & -0.08 \\
        & +2 & -0.67 & -0.54 & -0.57 & 0.02 & -0.02 & -0.04 & 0.04 & -0.07 & -0.07 \\
        & +4 & -0.59 & -0.46 & -0.50 & 0.01 & -0.01 & -0.03 & 0.03 & -0.07 & -0.07 \\
        & +6 & -0.54 & -0.40 & -0.45 & 0.01 & -0.01 & -0.01 & 0.01 & -0.07 & -0.07 \\
        & +8 & -0.51 & -0.36 & -0.42 & 0.01 & -0.01 & -0.01 & 0.01 & -0.07 & -0.07 \\ \hline \hline
\end{tabular*}
\end{table}

\begin{table}[!ht]
\caption{\label{tab:biaxial_strain-sia} Calculated SIA parameters for monolayer CrI$_3$ under biaxial strain. All values are represented in a Cartesian basis.}
\begin{tabular*}{\linewidth}{@{\extracolsep{\fill}} lccccccccc }
\hline\hline
Strain (\%) & -8 & -6 & -4 & -2 & 0 & +2 & +4 & +6 & +8 \\
$\mathcal{A}^{zz}_{ii}$  (meV) & -0.27 & -0.26 & -0.20 & -0.13 & -0.08 & -0.04 & 0.02 & 0.10 & 0.19 \\ \hline \hline
\end{tabular*}
\end{table}
\clearpage

\newpage
\section{Parameters for monolayer CrI$_\mathbf{3}$ under uniaxial strain}
\begin{table}[!ht]
\caption{\label{tab:uniaxial_strain_nn} Calculated exchange parameters for the NN pairs of monolayer CrI$_3$ under uniaxial strain. Corresponding (i-j) pairs are indicated in Fig. \ref{fig:pairs}. All values are given in a Cartesian basis.}
\begin{tabular*}{\linewidth}{@{\extracolsep{\fill}} crccccccccc }
\hline\hline
Pair & Strain & $\mathcal{J}^{xx}_{ij}$ & $\mathcal{J}^{yy}_{ij}$ & $\mathcal{J}^{zz}_{ij}$ & $ \mathcal{J}^{xy}_{ij}$ & $ \mathcal{J}^{yx}_{ij}$ & $\mathcal{J}^{xz}_{ij}$ & $\mathcal{J}^{zx}_{ij}$ & $ \mathcal{J}^{yz}_{ij}$ & $\mathcal{J}^{zy}_{ij}$ \\
(i-j)   & (\%) & (meV) & (meV) & (meV) & (meV) & (meV) & (meV) & (meV) & (meV) & (meV) \\ \hline
(1-2)   &     & & & & & & & & & \\
        & -8 & -2.05 & -1.22 & -1.84 &  0.07 &  0.07 &  0.03 &  0.03 & -0.56 & -0.56 \\
        & -6 & -2.80 & -1.88 & -2.52 &  0.05 &  0.05 &  0.03 &  0.03 & -0.59 & -0.59 \\
        & -4 & -3.38 & -2.39 & -3.05 &  0.03 &  0.03 &  0.02 &  0.02 & -0.61 & -0.61 \\
        & -2 & -3.90 & -2.85 & -3.53 &  0.01 &  0.01 &  0.02 &  0.02 & -0.63 & -0.63 \\
        & 0  & -4.34 & -3.24 & -3.96 &  0.00 &  0.00 &  0.00 &  0.00 & -0.65 & -0.65 \\
        & +2 & -4.65 & -3.47 & -4.23 & -0.03 & -0.03 & -0.01 & -0.01 & -0.66 & -0.66 \\
        & +4 & -4.92 & -3.68 & -4.47 & -0.05 & -0.05 & -0.03 & -0.03 & -0.67 & -0.67 \\
        & +6 & -5.15 & -3.85 & -4.68 & -0.06 & -0.06 & -0.05 & -0.05 & -0.68 & -0.68 \\
        & +8 & -5.36 & -3.99 & -4.88 & -0.07 & -0.07 & -0.07 & -0.07 & -0.70 & -0.70 \\ \hline
(2-3)   &     & & & & & & & & & \\
        & -8 & -3.66 & -4.12 & -4.02 & -0.34 & -0.34 & -0.44 & -0.44 & 0.29 & 0.29 \\
        & -6 & -3.62 & -4.09 & -3.99 & -0.37 & -0.37 & -0.48 & -0.48 & 0.31 & 0.31 \\
        & -4 & -3.57 & -4.05 & -3.95 & -0.41 & -0.41 & -0.51 & -0.51 & 0.32 & 0.32 \\
        & -2 & -3.52 & -4.03 & -3.93 & -0.44 & -0.44 & -0.53 & -0.53 & 0.32 & 0.32 \\
        & 0  & -3.53 & -4.07 & -3.97 & -0.50 & -0.50 & -0.57 & -0.57 & 0.33 & 0.33 \\
        & +2 & -3.44 & -4.02 & -3.95 & -0.50 & -0.50 & -0.58 & -0.58 & 0.32 & 0.32 \\
        & +4 & -3.41 & -4.05 & -3.99 & -0.54 & -0.54 & -0.60 & -0.60 & 0.31 & 0.31 \\
        & +6 & -3.41 & -4.10 & -4.05 & -0.57 & -0.57 & -0.61 & -0.61 & 0.30 & 0.30 \\
        & +8 & -3.44 & -4.18 & -4.14 & -0.61 & -0.61 & -0.62 & -0.62 & 0.29 & 0.29 \\ \hline
(2-7)   &     & & & & & & & & & \\
        & -8 &  1.99 &  1.49 &  1.43 & 0.42 & 0.42 & 0.59 & 0.59 & 0.28 & 0.28 \\
        & -6 &  0.08 & -0.44 & -0.44 & 0.43 & 0.43 & 0.58 & 0.58 & 0.29 & 0.29 \\
        & -4 & -1.42 & -1.96 & -1.90 & 0.45 & 0.45 & 0.57 & 0.57 & 0.30 & 0.30 \\
        & -2 & -2.62 & -3.19 & -3.08 & 0.46 & 0.46 & 0.56 & 0.56 & 0.31 & 0.31 \\
        & 0  & -3.53 & -4.07 & -3.97 & 0.50 & 0.50 & 0.57 & 0.57 & 0.33 & 0.33 \\
        & +2 & -4.03 & -4.63 & -4.47 & 0.50 & 0.50 & 0.56 & 0.56 & 0.34 & 0.34 \\
        & +4 & -4.37 & -5.00 & -4.80 & 0.52 & 0.52 & 0.56 & 0.56 & 0.36 & 0.36 \\
        & +6 & -4.59 & -5.24 & -5.02 & 0.55 & 0.55 & 0.57 & 0.57 & 0.38 & 0.38 \\
        & +8 & -4.74 & -5.40 & -5.17 & 0.58 & 0.58 & 0.58 & 0.58 & 0.40 & 0.40 \\ \hline \hline
\end{tabular*}
\end{table}
\clearpage

\newpage
\begin{table}[!t]
\caption{\label{tab:uniaxial_strain_nnn} Calculated exchange parameters for the NNN pairs of monolayer CrI$_3$ under uniaxial strain. Corresponding (i-j) pairs are indicated in Fig. \ref{fig:pairs}. All values are given in a Cartesian basis.}
\begin{tabular*}{\linewidth}{@{\extracolsep{\fill}} crccccccccc }
\hline\hline
Pair & Strain & $\mathcal{J}^{xx}_{ij}$ & $\mathcal{J}^{yy}_{ij}$ & $\mathcal{J}^{zz}_{ij}$ & $ \mathcal{J}^{xy}_{ij}$ & $ \mathcal{J}^{yx}_{ij}$ & $\mathcal{J}^{xz}_{ij}$ & $\mathcal{J}^{zx}_{ij}$ & $ \mathcal{J}^{yz}_{ij}$ & $\mathcal{J}^{zy}_{ij}$ \\
(i-j)   & (\%) & (meV) & (meV) & (meV) & (meV) & (meV) & (meV) & (meV) & (meV) & (meV) \\ \hline
(1-3)   &     & & & & & & & & & \\
        & -8 & -0.88 & -0.93 & -0.83 & 0.15 & -0.07 & 0.10 & -0.02 & -0.05 & 0.12 \\
        & -6 & -0.82 & -0.87 & -0.78 & 0.13 & -0.04 & 0.10 &  0.00 & -0.04 & 0.10 \\
        & -4 & -0.76 & -0.82 & -0.73 & 0.11 & -0.01 & 0.10 &  0.01 & -0.03 & 0.10 \\
        & -2 & -0.71 & -0.77 & -0.69 & 0.10 &  0.01 & 0.09 &  0.03 & -0.02 & 0.09 \\
        & 0  & -0.67 & -0.73 & -0.66 & 0.08 &  0.03 & 0.09 &  0.03 & -0.01 & 0.09 \\
        & +2 & -0.64 & -0.72 & -0.64 & 0.07 &  0.05 & 0.09 &  0.05 & -0.01 & 0.09 \\
        & +4 & -0.63 & -0.71 & -0.64 & 0.06 &  0.07 & 0.09 &  0.06 &  0.00 & 0.09 \\
        & +6 & -0.62 & -0.71 & -0.64 & 0.06 &  0.09 & 0.09 &  0.07 &  0.00 & 0.09 \\
        & +8 & -0.61 & -0.71 & -0.65 & 0.05 &  0.10 & 0.09 &  0.08 &  0.01 & 0.09 \\ \hline
(1-7)   &     & & & & & & & & & \\
        & -8 & -1.07 & -1.14 & -1.08 & -0.20 & 0.05  & -0.09 & -0.11 &  0.04 & 0.07 \\
        & -6 & -0.94 & -1.01 & -0.94 & -0.16 & 0.02  & -0.09 & -0.09 &  0.03 & 0.07 \\
        & -4 & -0.84 & -0.91 & -0.83 & -0.13 & 0.01  & -0.09 & -0.07 &  0.02 & 0.07 \\
        & -2 & -0.75 & -0.81 & -0.74 & -0.10 & -0.01 & -0.09 & -0.06 &  0.01 & 0.08 \\
        & 0  & -0.67 & -0.73 & -0.66 & -0.08 & -0.03 & -0.09 & -0.03 & -0.01 & 0.09 \\
        & +2 & -0.61 & -0.67 & -0.60 & -0.06 & -0.04 & -0.09 & -0.03 & -0.01 & 0.08 \\
        & +4 & -0.55 & -0.61 & -0.54 & -0.05 & -0.05 & -0.09 & -0.02 & -0.02 & 0.08 \\
        & +6 & -0.50 & -0.55 & -0.50 & -0.04 & -0.06 & -0.09 & -0.01 & -0.02 & 0.09 \\
        & +8 & -0.47 & -0.52 & -0.47 & -0.03 & -0.07 & -0.09 &  0.00 & -0.03 & 0.09 \\ \hline
(9-13)  &     & & & & & & & & & \\
        & -8 & -1.12 & -0.99 & -1.00 & 0.13 & -0.13 & -0.10 & 0.06 & -0.04 & -0.12 \\
        & -6 & -1.00 & -0.87 & -0.88 & 0.10 & -0.09 & -0.08 & 0.05 & -0.05 & -0.10 \\
        & -4 & -0.90 & -0.78 & -0.79 & 0.06 & -0.07 & -0.07 & 0.05 & -0.06 & -0.09 \\
        & -2 & -0.82 & -0.70 & -0.71 & 0.04 & -0.05 & -0.06 & 0.05 & -0.07 & -0.08 \\
        & 0  & -0.76 & -0.64 & -0.66 & 0.03 & -0.03 & -0.05 & 0.05 & -0.08 & -0.08 \\
        & +2 & -0.72 & -0.59 & -0.62 & 0.01 & -0.02 & -0.05 & 0.06 & -0.09 & -0.06 \\
        & +4 & -0.69 & -0.56 & -0.59 & -0.01 & -0.01 & -0.05 & 0.07 & -0.10 & -0.06 \\
        & +6 & -0.67 & -0.54 & -0.58 & -0.02 & -0.01 & -0.06 & 0.07 & -0.10 & -0.05 \\
        & +8 & -0.66 & -0.52 & -0.57 & -0.03 &  0.00 & -0.06 & 0.08 & -0.11 & -0.04 \\ \hline \hline
\end{tabular*}
\end{table}

\begin{table}[!t]
\caption{\label{tab:uniaxial_strain-sia} Calculated SIA parameters for monolayer CrI$_3$ under uniaxial strain. All values are represented in a Cartesian basis. Notice that $\mathcal{A}^{xy}_{ii} = \mathcal{A}^{yx}_{ii}$, $\mathcal{A}^{xz}_{ii} = \mathcal{A}^{zx}_{ii}$, and $\mathcal{A}^{yz}_{ii} = \mathcal{A}^{zy}_{ii}$ since the SIA matrix is symmetric. As detailed in Ref. \cite{sabani2020}, one needs to calculate only five terms to capture the full effect of the SIA \cite{remark-sia}.}
\begin{tabular*}{\linewidth}{@{\extracolsep{\fill}} lccccccccc }
\hline \hline
Strain (\%) & -8 & -6 & -4 & -2 & 0 & +2 & +4 & +6 & +8 \\ \hline
$\mathcal{A}^{yy}_{ii}-\mathcal{A}^{xx}_{ii}$ (meV) & -0.11 & -0.10 & -0.07 & -0.03 &  0.00 &  0.05 &  0.07 & 0.10 & 0.11 \\ 
$\mathcal{A}^{zz}_{ii}-\mathcal{A}^{xx}_{ii}$ (meV) & -0.26 & -0.22 & -0.16 & -0.12 & -0.08 & -0.05 & -0.02 & 0.01 & 0.03 \\ 
$\mathcal{A}^{xy}_{ii}$ (meV) & -0.02 & -0.01 & -0.01 & -0.01 &  0.00 &  0.00 &  0.01 &  0.01 & 0.03\\ 
$\mathcal{A}^{xz}_{ii}$ (meV) &  0.25 &  0.16 & 0.10 &  0.04 &  0.00 & -0.03 & -0.03 & -0.03 & -0.03 \\ 
$\mathcal{A}^{yz}_{ii}$ (meV) &  0.07 &  0.04 & 0.03 &  0.01 &  0.00 & -0.01 & -0.01 & -0.02 & -0.01 \\ \hline \hline
\end{tabular*}
\end{table}
\clearpage

\newpage
\section{Parameters for monolayer CrI$_\mathbf{3}$ under electric field}
\begin{table}[!ht]
\caption{\label{tab:efield} Calculated exchange parameters for monolayer CrI$_3$ under uniform applied electric field. Corresponding (i-j) pairs are indicated in Fig. \ref{fig:pairs}. All values are given in a Cartesian basis.}
\begin{tabular*}{\linewidth}{@{\extracolsep{\fill}} ccccccccccc }
\hline\hline
Pair & Field & $\mathcal{J}^{xx}_{ij}$ & $\mathcal{J}^{yy}_{ij}$ & $\mathcal{J}^{zz}_{ij}$ & $ \mathcal{J}^{xy}_{ij}$ & $ \mathcal{J}^{yx}_{ij}$ & $\mathcal{J}^{xz}_{ij}$ & $\mathcal{J}^{zx}_{ij}$ & $ \mathcal{J}^{yz}_{ij}$ & $\mathcal{J}^{zy}_{ij}$ \\
(i-j)   & (eV/\AA{}) & (meV) & (meV) & (meV) & (meV) & (meV) & (meV) & (meV) & (meV) & (meV) \\ \hline
(1-2)   &     & & & & & & & & & \\
        & 0.0 & -4.34 & -3.24 & -3.96 &  0.00 &  0.00 &  0.00 & 0.00 & -0.65 & -0.65 \\
        & 0.2 & -4.35 & -3.23 & -3.95 &  0.00 & -0.03 & -0.07 & 0.07 & -0.65 & -0.65 \\
        & 0.4 & -4.35 & -3.23 & -3.95 &  0.01 & -0.04 & -0.14 & 0.15 & -0.65 & -0.65 \\
        & 0.6 & -4.35 & -3.23 & -3.95 &  0.02 & -0.05 & -0.21 & 0.22 & -0.66 & -0.66 \\ \hline
(2-3)   &     & & & & & & & & & \\
        & 0.0 & -3.53 & -4.07 & -3.97 & -0.50 & -0.50 & -0.57 & -0.57 & 0.33 & 0.33 \\
        & 0.2 & -3.52 & -4.08 & -3.96 & -0.50 & -0.48 & -0.61 & -0.54 & 0.27 & 0.39 \\
        & 0.4 & -3.52 & -4.08 & -3.96 & -0.52 & -0.47 & -0.64 & -0.50 & 0.20 & 0.45 \\
        & 0.6 & -3.52 & -4.08 & -3.96 & -0.53 & -0.46 & -0.68 & -0.47 & 0.15 & 0.52 \\ \hline
(2-7)   &     & & & & & & & & & \\
        & 0.0 & -3.53 & -4.07 & -3.97 & 0.50 & 0.50 & 0.57 & 0.57 & 0.33 & 0.33 \\
        & 0.2 & -3.52 & -4.08 & -3.96 & 0.49 & 0.52 & 0.53 & 0.60 & 0.38 & 0.26 \\
        & 0.4 & -3.52 & -4.08 & -3.96 & 0.48 & 0.53 & 0.50 & 0.64 & 0.45 & 0.20 \\
        & 0.6 & -3.52 & -4.08 & -3.96 & 0.47 & 0.54 & 0.47 & 0.68 & 0.51 & 0.14 \\ \hline
(1-3)   &     & & & & & & & & & \\
        & 0.0 & -0.67 & -0.73 & -0.66 & 0.08 & 0.03 & 0.09 & 0.03 & -0.01 & 0.09 \\
        & 0.2 & -0.67 & -0.73 & -0.66 & 0.08 & 0.03 & 0.10 & 0.03 & -0.01 & 0.08 \\
        & 0.4 & -0.68 & -0.73 & -0.66 & 0.08 & 0.03 & 0.10 & 0.03 & -0.01 & 0.08 \\
        & 0.6 & -0.68 & -0.74 & -0.67 & 0.08 & 0.03 & 0.11 & 0.02 &  0.00 & 0.08 \\ \hline
(1-7)   &     & & & & & & & & & \\
        & 0.0 & -0.67 & -0.73 & -0.66 & -0.08 & -0.03 & -0.09 & -0.03 & -0.01 & 0.09 \\
        & 0.2 & -0.67 & -0.73 & -0.66 & -0.08 & -0.03 & -0.09 & -0.04 & -0.01 & 0.09 \\
        & 0.4 & -0.69 & -0.76 & -0.66 & -0.08 & -0.03 & -0.09 & -0.04 & -0.01 & 0.09 \\
        & 0.6 & -0.68 & -0.74 & -0.67 & -0.08 & -0.03 & -0.08 & -0.04 & -0.02 & 0.10 \\ \hline
(9-13)  &     & & & & & & & & & \\
        & 0.0 & -0.76 & -0.64 & -0.66 & 0.03 & -0.03 & -0.05 & 0.05 & -0.08 & -0.08 \\
        & 0.2 & -0.76 & -0.64 & -0.66 & 0.03 & -0.03 & -0.05 & 0.05 & -0.08 & -0.07 \\
        & 0.4 & -0.76 & -0.64 & -0.66 & 0.03 & -0.03 & -0.05 & 0.05 & -0.09 & -0.07 \\
        & 0.6 & -0.77 & -0.65 & -0.67 & 0.03 & -0.03 & -0.05 & 0.05 & -0.09 & -0.06 \\ \hline \hline
\end{tabular*}
\end{table}

\begin{table}[!ht]
\caption{\label{tab:efield-sia} Calculated SIA parameters for monolayer CrI$_3$ under uniform applied electric field. All values are represented in a Cartesian basis.}
\begin{tabular}{cc}
\hline\hline
Field    & \qquad $\mathcal{A}^{zz}_{ii}$  \\ 
(eV/\AA{}) & \qquad (meV) \\ \hline
0.0  & \qquad -0.08 \\
0.2  & \qquad -0.08 \\
0.4  & \qquad -0.08 \\
0.6  & \qquad -0.08 \\ \hline \hline
\end{tabular}
\end{table}